# Analysis of a Symmetry leading to an Inertial Range Similarity Theory for Isotropic Turbulence.

Mogens V Melander

Department of Mathematics

Southern Methodist University, Dallas TX 75275-0156

We present a theoretical attack on the classical problem of intermittency and anomalous scaling in turbulence. Our focus is on an ideal situation: high Reynolds number isotropic turbulence driven by steady large scale forcing. Moreover, the fluid is incompressible and no confining boundaries are present. We start from a good set of basis functions for the velocity field. These are real and divergence-free. To each wave-vector $\vec{k}$ in Fourier space there is one pair of basis functions with respectively left and right-handed polarity. Isotropy makes all $\vec{k}$ on the shell of constant $\left|\vec{k}\right|$ statistically equivalent. Consequently, the coefficients , $\chi^{+}$ and $\chi^{-}$ , to the basis functions in that shell become two random variables whose joint pdf describes the statistics at scale $\ell = 2\pi / k$ . Moreover, $\left(\chi^{+}\right)^{2} + \left(\chi^{-}\right)^{2}$ becomes a random variable for the energy. Switching to polar coordinates, the joint pdf expands in azimuthal modes. We focus on the axisymmetric mode which is itself a pdf and characterized by it radial profile $P_{0}(r;\ell)$ . Observations from both shell model and DNS data indicate that (1) the moments of $P_{0}(r;\ell)$ scale as power laws in $\ell$ , and (2) the profile obeys an affine symmetry $P_{0}(r;\ell) = C(\ell) f\left(\dfrac{\ln r - \mu(\ell)}{\sigma(\ell)}\right)$ . We raise the question: What statistics agree with both observation? The answer is pleasing. We find the functions $f, \mu, \sigma$ and $C$ analytically in terms of a few constants. Moreover, we obtain closed form expressions for both scaling exponents and coefficients in the power laws. A virtual origin also emerges as an intrinsic length scale $\ell_{o}$ for the inertial range.

## 1. Introduction

Symmetry is our leitmotif for this paper. Symmetry has a long history in theoretical physics and has led to many key results. It is possible that symmetry may also resolve classical problems in turbulence. There, symmetry shows up in statistical properties. In the case of turbulence driven by steady forcing, for example, time-independence occurs only in a statistical sense: averages over an ensemble of flow realizations are stationary, but individual realizations vary chaotically with time. Similarly, Frisch (1995) points out that far away from boundaries, turbulence may restore the symmetries of Navier Stokes equations statistically. Turbulence may also develop new symmetries in addition to those native to Navier Stokes equations (i.e., Frisch 1995 p 17). To see how this is possible, let us think of Navier Stokes equations as a dynamical system with an attractor describing equilibrium turbulence. The attractor occupies a small subset of phase space. Conceivably, this subset can have a self-similar structure, for example when sliced by scale. This way, new symmetries can emerge when we restrict Navier Stokes equations to the subset. We can thus speak of "attractor symmetries." Let us give a simple illustration of this concept. Imagine a dynamical system

$$dx_1/dt = f_1(x_1, x_2), \qquad dx_2/dt = f_2(x_1, x_2) \qquad (1.1)$$

with a phase portrait as shown in Figure 1 where the circle $x_1^2 + x_2^2 = 1$ is a globally attracting limit cycle. Moreover, suppose (1.1) restricted to the attractor reads $dx_1/dt = -x_2$, $dx_2/dt = x_1$. In that scenario, we have rotational symmetry on the attractor, but not elsewhere. So we have an attractor specific symmetry. Suppose we consider a large ensemble of solutions to (1.1) then the symmetry emerges statistically in the long time limit. For example, we will see it in the moments: $\left\langle \left(x_1^2 + x_2^2\right)^{p/2}\right\rangle \to 1$ as $t \to \infty$, where "$\langle \ \rangle$" denotes ensemble average. Note, these moments correspond to the large time asymptotic pdf of our ensemble: $\delta(1 - x_1^2 - x_2^2)/2\pi$.

While vastly more complicated than (1), 3D turbulence also has attractor specific symmetries. One example is the dissipation range. Away from boundaries, the dissipation range is generally believed to be universal when expressed in the proper units, i.e., the Kolmogorov scale $\eta_{Kol}$, and dissipation $\varepsilon$. Another example, is the inertial range at high Reynolds numbers where "scaling exponents" for various structure functions are thought to be universal; e.g., Frisch (1995).

Symmetry performs best under idealized circumstances. Problems under that rubric are precisely those one would expect to have been solved eons ago. In other mature fields of science such problems are textbook material and their solutions constitute established indisputable knowledge. That is not the case for turbulence. The unresolved nature of even the most fundamental and simple problems justifies turbulence reputation as "the last great problem in classical physics". Although the origin of this quote is unclear, the quote is by no means trite. The fact that basic problems remain unsolved is of greater theoretical significance than the complexity arising from turbulent flows having a huge number of degrees of freedom. Moreover, the lack of basic theoretical results impedes turbulence modeling.

One unresolved problem involves steady large scale forcing of isotropic 3D turbulence at a high Reynolds number in an incompressible fluid with constant properties. In absence of boundaries, we have all possible simplifications in our favor. Naturally, this problem has a long history. It is thus well known that a universal "inertial (sub)range" develops. This range covers length scales where inertial forces rule. These scales are much smaller than the forcing scale, but also much larger than the dissipation scale. In the large Reynolds number limit, the inertial range becomes infinitely wide. To analyze the statistics in the inertial range, structure functions of velocity differences between two points are traditionally used. Say the points are $P$ and $Q$, then velocity increments parallel and perpendicular to $\overrightarrow{PQ}$ are of interest. Because of isotropy the statistics of these differences depend only on the separation distance $\ell = |PQ|$ between the two points. It is known that the statistical moments of the velocity differences obey power laws, i.e.,

$$S_{\parallel p}(\ell) \equiv \left\langle \left( \delta v_{\parallel}(\ell) \right)^p \right\rangle = C_{p\parallel} \ell^{\zeta_{p\parallel}} \quad \text{and} \quad S_{\perp p}(\ell) \equiv \left\langle \left( \delta v_{\perp}(\ell) \right)^p \right\rangle = C_{p\perp} \ell^{\zeta_{p\perp}} \tag{1.3}$$

Kolmogorov's four-fifth law (e.g., Frisch 1995 p 76) states

$$\left\langle \left( \delta v_{\parallel}(\ell) \right)^3 \right\rangle = -4/5 \varepsilon \ell \tag{1.4}$$

thereby providing both $C_{3\parallel}$ and $\zeta_{3\parallel}$. By definition, we also have $C_{0\parallel} = 0$ and $\zeta_{0\parallel} = 0$. For other, in particular high, values of $p$, the picture is less clear.

Relying on the idea that the physics is the same everywhere in the inertial range, Kolmogorov (1941) used (1.4) to suggest scaling laws for all moments:

$$S_{\parallel p}(\ell) = \tilde{C}_{p\parallel} \varepsilon^{p/3} \ell^{p/3} \tag{1.5}$$

with universal constants $\tilde{C}_{p\parallel}$. In that scenario, dimensionless ratios of moments such as $S_6 / S_3^2$ are also universal constants. Kolmogorov's suggestion is an attractor specific symmetry stating that the pdf $\phi_\parallel$ for $\delta v_\parallel (\ell)$ should vary self-similarly with scale according to the formula

$$\phi_\parallel (x; \ell) = S_{\parallel 2}^{-1/2} (\ell) \phi_\parallel (x / S_{\parallel 2}^{1/2} (\ell); 1) \; . \tag{1.6}$$

This symmetry is known as statistical self-similarity, and occurs, for example, in Brownian motion (e.g., Embrechts & Maejima 2002). The linear scaling exponents $\zeta_p = p/3$ are called "K41 scaling". We note the exponents *p/3* do not specify the shape of the pdf $\phi_\parallel$. That also requires the coefficients $\tilde{C}_{\parallel p}$. Even then, knowing all the integer order moments may not be enough to uniquely specify $\phi_\parallel$ because of "Haussdorff's moment problem" ; see Koerner (1988) p 21.

A large body of evidence refutes K41 scaling in the inertial range of 3D-turbulence; see e.g., Frisch (1995), Benzi et al (1995), Sreenivasan & Antonia (1997), Katul, et al (2001). That is, the scaling exponent is a nonlinear function of *p*. Equivalently, the pdf's do not collapse according to (1.6). As shown in Figure 2, the tails of the pdf flare out and thicken with decreasing $\ell$ even after normalizing by the standard deviation. That means fluctuations of many standard deviations become increasingly more likely with decreasing $\ell$. Such deviation from statistical self-similarity constitutes "intermittency" as defined by Frisch (1995) p 121. In his definition, the issue is not much the pdf deviates from a Gaussian, but that the dependence on $\ell$ deviates from (1.6). Correspondingly, the structure functions are said to scale "anomalously" when $\zeta_p / p \neq Const$. Interestingly, the longitudinal and transverse scaling exponents (i.e., $\zeta_{\parallel p}$ and $\zeta_{\perp p}$) are different, but seem related through a simple formula; see Siefert & Peinke (2004, 2006). Also, in contrast to $\zeta_{\parallel p}$, $\zeta_{\perp p}$ can be obtained for non-integer orders.

A key difference between K41 and anomalous scaling is the existence of an intrinsic scale $\ell_o$. Statistical self-similarly (1.6) does not provide one, so a reference scale is supplied from elsewhere. K41 usually uses the integral scale, which depends on large eddies outside the inertial range. Being larger than the scales in the inertial range, the integral scale says something about where the inertial range begins. Similarly, the dissipation scale, $\eta_{Kol}$, says something about where the inertial range stops. Neither the beginning nor the end can be found directly from (1.6) alone. In principle, $\ell$ could run from zero to infinity in (1.6). For anomalous scaling, however, $\ell$ cannot run to zero without violating inertial range scaling laws. We illustrate this point using longitudinal structure functions. For any random variable $R$, we have

$$0 \leq \left\langle \left( R^{2p} - \left\langle R^{2p} \right\rangle \right)^2 \right\rangle = \left\langle R^{4p} \right\rangle - \left\langle R^{2p} \right\rangle^2 = S_{4p} - S_{2p}^2 \,. \tag{1.7}$$

Suppose $R = \delta v_{\parallel}$, then the power laws (1.3) can then be substituted in (1.7) to give

$$0 \leq C_{4p\parallel} \ell^{\zeta_{4p\parallel}} - C_{2p\parallel}^2 \ell^{2\zeta_{2p}} = C_{4p\parallel} \ell^{\zeta_{4p\parallel}} \left( 1 - \frac{C_{2p\parallel}^2}{C_{4p\parallel}} \ell^{2\zeta_{2p\parallel} - \zeta_{4p\parallel}} \right) \,. \tag{1.8}$$

Here, the first factor is always positive, but the last factor will change sign at some $\ell$ when $2\zeta_{2p\parallel} \neq \zeta_{4p\parallel}$. The inequality then states that the second factor must not be negative. Thus, anomalous scaling laws can never be extended to all $\ell$ like K41. Whether the second factor is positive on the infrared or the ultraviolet side depends on the sign of the exponent $2\zeta_{2p\parallel} - \zeta_{4p\parallel}$. It known that $\zeta_p$ is a concave down function of $p$ in the inertial range (Frisch 1995 p 133) so that $2\zeta_{2p\parallel} - \zeta_{4p\parallel} \geq 0$. For example, reported experimental values are $\zeta_{2\parallel} = 0.70$ and $\zeta_{4\parallel} = 1.28$ (Frisch 1995 p 131, Benzi et al 1995). Consequently, (1.8) shows that the anomalous scaling laws terminate on the infrared side at some $\ell_o > 0$ which is then an intrinsic length scale.

Let us reconsider the statement that $\zeta_p$ must be concave down in the inertial range. It is indeed an exact result, if properly qualified. Schwarz's inequality for any random variable $R$ leads to

$$\left\langle \left| R \right|^p \right\rangle^2 \leq \left\langle \left| R \right|^{p+h} \right\rangle \left\langle \left| R \right|^{p-h} \right\rangle, \quad 0 \leq h \leq p ; \tag{1.9}$$

see Feller (1966), Vol 2, p 153. Say we have a positive random variable whose structure functions obey power laws: $S_p(\ell) = C_p \ell^{\zeta_p}$. Substituting these power laws into (1.9) and rearranging yields:

$$0 < \frac{C_p^2}{C_{p+h} C_{p-h}} \leq \ell^{\zeta_{p+h} + \zeta_{p-h} - 2\zeta_p} \quad . \tag{1.10}$$

Here we notice that $q \equiv \left( \zeta_{p+h} + \zeta_{p-h} - 2\zeta_p \right)/h^2$ is the difference quotient for $\zeta_p''$, so $\zeta_p$ is concave up when $q$ is positive and concave down when $q$ is negative. The inequality (1.10) can not be satisfied for all $\ell$ when $q \neq 0$. In the ultraviolet limit, $\ell \to 0$, our inequality is satisfied when $q < 0$ and hence $\zeta_p$ must be concave down. In the infrared limit, $\ell \to \infty$, the inequality requires $q > 0$ and hence $\zeta_p$ must be concave up. The classical inertial range sits between a large forcing scale and an arbitrary small dissipation scale. That is the ultraviolet limit, where $\zeta_p$ must be concave down. On the other hand, were we to consider small scale forcing and a cascade from small to large scales (e.g., Gibson 1996) then the infrared limit would come in play and $\zeta_p$ would be concave up.

Besides concavity, the literature (e.g., Frisch 1995) mentions to two other constraints on $\zeta_p$. One states that $\zeta_p$ must be an increasing function of $p$. The other is "Novikov's inequality" stating that $\zeta_p = p/3 + \tau_{p/3}$ where $\tau_p + 3p \geq 0$ for $p \geq 0$ and $\tau_p + 3p \leq 0$ for $p \leq 0$ (see Frisch 1995 p 172, Novikov 1970). We dispute both constraints. Let us clear up these issues now so that they will not be of concern later.

The argument leading to $\zeta_p$ being an increasing function rests on the assumption of a largest velocity. Without this assumption, supersonic velocities could be present and violate the use of the incompressible Navier Stokes equations. The counter argument is that the incompressible equations constitute a complete mathematical framework. Once we decide to work within that framework, we must accept its inherent statistics, whatever they are. No additional "modeling constraint" such as a largest velocity may be imposed. The fact that the incompressible framework gives an incorrect description of real physical flows at very high speeds is a different matter. If the incompressible statistics show that the velocity can become unbounded then we must accept that compressible corrections are required to describe real flows.

To arrive at the Novikov inequality, one considers structure functions for the dissipation at scale $\ell$. The corresponding scaling exponents are called $\tau_p$. It is believed that $\zeta_p - p/3 = \tau_{p/3}$ is the correction to the K41 scaling. We do not dispute this relation provided that the structure functions for the dissipation are defined appropriately. The latter point comes down to how the "dissipation at scale $\ell$" should be defined. In our view, an appropriate definition involves transforming the physical space dissipation (i.e., $\varepsilon = 2\nu S_{ij}S_{ij}$, where $S_{ij}$ is the strain rate tensor) to Fourier space and filtering according to scale, i.e., $\ell = 2\pi / |\vec{k}|$. In contrast, the common usage defines the dissipation at scale $\ell$ as the average of $2\nu S_{ij}S_{ij}$ over a sphere of radius $\ell$. Correspondingly, the $p$'th order structure function is the average of $\left(2\nu S_{ij}S_{ij}\right)^p$ over a sphere of radius $\ell$. With that definition, Novikov's inequality follows from the mathematical fact that the integral of the positive function $\left(2\nu S_{ij}S_{ij}\right)^p$ is a non-decreasing function of the radius of the sphere (see Frisch 1995, p 173). Once the usual definition of the dissipation structure function is accepted, Novikovs inequality follows. Our objection is that this traditional definition also includes all scales smaller than $\ell$. If we use Fourier space to filter the dissipation according to scale then Novikov's inequality disappears.

A large variety of models exist for $\zeta_p$, a small partial list includes Kolmogov (1962), Parisi & Frisch (1985), She & Leveque (1994), Lundgreen (2006). Generally, most models agree with observations so long $p$ is moderate. In particular, the "log-Poisson" model (She & Leveque 1994) is famous for faithfully reproducing experimentally observed exponents (e.g., Benzi et al 1995). For large values of $p$, the models produce widely differing results. It would then seem that experiments and direct numerical simulations should easily be able to reject most models on the basis of high order exponents. Unfortunately, the reality is different because high orders exponents are difficult to obtain accurately from data, e.g., Wit (2004). One of the obstacles is that one has to work with a finite sample. The moments of a finite sample can be expected to agree with those of the continuous pdf when the order is low, but at high order the two kinds of moments are certain to disagree, possibly catastrophically. In particular, we point out that a finite sample can be viewed as a pdf with compact support. Moments of unreasonably high orders will act accordingly. In fact, they can act in such a way as to support the idea of a monotone $\zeta_p$. For these reasons, we must view high order exponents with skepticism. Many models (e.g., Kolmogorov 1962) are often thought to misrepresent high order exponents on the grounds that (1) $\zeta_p$ is not monotone, (2) Novikov's inequality is violated, or (3) there is strong disagreement with experimental data. In light of our arguments, no model can be rejected for those reasons.

This is paper is motivated by the idea of symmetry. From a classical point of view, we should expect symmetry in the inertial range because the same physics, inertia, rule at all scales. Thus, we search for a global scaling invariance for the inertial range, such that the full statistics at one scale, $\ell$, suffice to generate the full statistics at all other scales through a similarity formula. We approach this symmetry issue with a new set of variables for describing turbulence. These variables are introduced in Section 2. In contrast to most of the literature, we do not work with physical space velocity increments, but with amplitudes in Fourier space. Correspondingly, our scale $\ell$ is defined three dimensionally in Fourier space and not as a separation between points in physical space. Our variables suggest symmetry and where to look for it. We observe the expected symmetry in turbulence data. The data comes from both a shell model and DNS. Moreover, the data does exhibit intermittency in the usual sense; see Figure 2. Justified by observations, Section 2 states the symmetry as a hypothesis (H1) and augments it with the traditional power law assumption (H2). Together, H1 and H2 are taken as exact properties of turbulence. The key question is this: under what circumstances can both H1 and H2 be satisfied. We analyze this question in Section 3. The result is a functional equation for intermittency with K41-scaling as the trivial solution. Section 4 solves this equation completely, thereby providing both all scaling exponents $\zeta_p$ and all coefficients $C_p$. The intrinsic length scale $\ell_o$ for anomalous scaling also shows up. Section 5 wraps up the analysis by examining properties of the solution and finding the pdf corresponding to $\zeta_p$ and $C_p$. Section 6 presents the complete set of formulas for the structure functions and the underlying pdf. Section 6 also concludes the paper with a discussion of why the choice of variables matters.

## 2. The Setup: Variables, Observations and Assumptions

Navier Stokes equations for an incompressible constant property fluid describe the evolution of a solenoidal (i.e., divergence free) vector field. Two real scalars suffice to describe such a field. For example, two Clebsch potentials can do the job, e.g., $\vec{u} = \varphi \nabla \psi$. However, we want our scalars to describe the velocity in a way that reflects the left-right symmetry in Navier Stokes equations. This we can accomplish by using the "complex helical waves decomposition"; see Lesieur (1990). It uses eigenfunctions of the curl operator as basis functions. Those separate into two kinds, left and right handed, according to the sign of the corresponding eigenvalue. The decomposition is straightforward to perform in Fourier space and provides us with two amplitudes, $u^-\left(\vec{k}\right)$ and $u^+\left(\vec{k}\right)$, for each wave vector $\vec{k}$. The amplitudes are related to the Fourier transformed velocity $\hat{\vec{u}}(\vec{k})$ as follows (Lesieur 1990 p 99): $2u^\pm(\vec{k}) = \hat{\vec{u}}(\vec{k}) \cdot (\vec{b}(\vec{k}) \pm i\vec{a}(\vec{k}))$, where $\vec{a}(\vec{k})$ and $\vec{b}(\vec{k})$ are, respectively, azimuthal and longitudinal unit vectors on the sphere of constant $\left|\vec{k}\right|$. The amplitudes are complex, but obey conjugate symmetry, e.g., $u^-\left(-\vec{k}\right) = u^{-*}\left(\vec{k}\right)$. This symmetry results from the velocity field being real in physical space. Conjugate symmetry allows us to introduce two real variables, $\chi^- = \mathrm{Re}(u^-) + \mathrm{Im}(u^-)$ and $\chi^+ = \mathrm{Re}(u^+) + \mathrm{Im}(u^+)$, so that

$$u^\pm\left(\vec{k}\right) = \frac{1}{2}\left(\chi^\pm\left(\vec{k}\right) + \chi^\pm\left(-\vec{k}\right)\right) + \frac{i}{2}\left(\chi^\pm\left(\vec{k}\right) - \chi^\pm\left(-\vec{k}\right)\right). \tag{2.1}$$

Hereby, conjugate symmetry is identically satisfied. Thus, we have no constraints on $\chi^-$ and $\chi^+$. Corresponding to $\chi^-$ and $\chi^+$ we have real basis functions. These we obtain by introducing $\chi^-$ and $\chi^+$ into the complex helical waves decomposition. The result is:

$$\vec{u}(\vec{r}) = \frac{1}{\sqrt{2}}\int_{\vec{k}}\chi^+(\vec{k})\left[\cos(\vec{k}\cdot\vec{r} - \pi/4)\vec{a}(\vec{k}) + \cos(\vec{k}\cdot\vec{r} + \pi/4)\vec{b}(\vec{k})\right]d\vec{k}$$

$$+ \frac{1}{\sqrt{2}}\int_{\vec{k}}\chi^-(\vec{k})\left[\cos(\vec{k}\cdot\vec{r} - \pi/4)\vec{b}(\vec{k}) + \cos(\vec{k}\cdot\vec{r} + \pi/4)\vec{a}(\vec{k})\right]d\vec{k}. \tag{2.2}$$

We notice that the third unit vector, $\vec{c}(k) = \vec{k}/\left|\vec{k}\right|$, on the sphere of constant $\left|\vec{k}\right|$ is absent from (2.2). That is how we can immediately tell that (2.2) is divergence free;

$i\vec{k} \cdot \hat{\vec{u}}(\vec{k}) = 0$, or $\vec{c}(\vec{k}) \cdot \hat{\vec{u}}(\vec{k}) = 0$, is namely the Fourier transformed incompressibility condition. The basis vectors in (2.2) are orthonormal in the usual Fourier sense, so we have

$$\int_{\mathbb{R}^3} \left| \vec{u}(\vec{r}) \right|^2 d\vec{r} = \frac{1}{2} \int_{\mathbb{R}^3} \left( \chi^+(\vec{k}) \right)^2 + \left( \chi^-(\vec{k}) \right)^2 d\vec{k} \ . \qquad (2.3)$$

Our two scalars thus build the energy from their square sum in a Cartesian fashion.

Our scalars, $\chi^-(\vec{k})$ and $\chi^+(\vec{k})$, provide the foundation for our investigation of turbulence. These scalar variables have very nice properties. We emphasize four of these. First, our variables are unconstrained. This means that regardless of their values formula (2.2) always describes a real solenoidal vector field. Second, the variables provide a complete representation of the flow field. This means that the Navier Stokes equations can be expressed as a dynamical system in $\chi^-(\vec{k})$ and $\chi^+(\vec{k})$ without using any modeling. Third, the variables enter symmetrically in expressions for the energy, enstrophy, etc; see (2.3). That is because the variables are designed to describe the left-right symmetry native to Navier Stokes equations. Finally, our variables are directly associated with the notion of scale, namely in the Fourier sense: $\ell = 2\pi / \left| \vec{k} \right|$. In the light of these properties, we can use $\chi^-(\vec{k})$ and $\chi^+(\vec{k})$ for practically anything, including a statistical description of turbulence. Focusing specifically on isotropic turbulence, we know that all $\vec{k}$ with the same magnitude are the statistically equivalent. At each scale, $\ell = 2\pi / \left| \vec{k} \right|$, the collection of all $(\chi^-, \chi^+)$ therefore generates a two dimensional pdf. Thinking in terms of dynamical systems, this pdf gives a statistical description of that attractor in phase space which represents equilibrium turbulence. Within this description, $\chi^-$ and $\chi^+$ serve as two random variables. To emphasize the Cartesian nature expressed by (2.3) we call these two random variables $X$ and $Y$. Considering an ensemble of realizations of the same turbulent flow we let $X = \chi^-(\vec{k}_o)$ and $Y = \chi^+(\vec{k}_o)$ where $\vec{k}_o$ is an arbitrary fixed wave vector with magnitude $\left|\vec{k}_o\right| = 2\pi/\ell$. (If we invoke ergodicity, a single realization suffices. $\vec{k}_o$ would then be a random vector on the shell $\left|\vec{k}_o\right| = 2\pi/\ell$.)

The joint pdf $\mathcal{J}$ of our two random variables describes the statistics at scale $\ell$:

$$\mathcal{J}(x,y,\ell)dxdy \equiv \Pr\{x < X < x+dx, y < Y < y+dy\}. \qquad (2.4)$$

On account of (2.3), we can also construct random variables for energy, enstrophy, etc at scale $\ell$ from $X^2 + Y^2$. (Helicity involves $X^2 - Y^2$.) The graph of $\mathcal{J}$ has the shape of a mountain located near $(x,y) = (0,0)$, because $\mathcal{J} \to 0$ as $x^2 + y^2 \to \infty$. For this reason, we switch to polar coordinates $(x,y) = r(\cos\theta, \sin\theta)$ and automatically obtain an azimuthal expansion

$$\mathcal{J}(x,y;\ell) = P_0(r,\ell) + \cos(\theta - \theta_1(\ell))P_1(r,\ell) + \cos(2(\theta - \theta_2(\ell)))P_2(r,\ell) + ... \quad (2.5)$$

where $\theta_1, \theta_2, ...$ are phase-constants. This expansion is valid in general for isotropic turbulence in equilibrium. In this paper, we focus on the axisymmetric component of $\mathcal{J}$:

$$\mathcal{J}_o(x,y,\ell) = P_0(\sqrt{x^2 + y^2}, \ell). \qquad (2.6)$$

Averaging (2.5) over $\theta$ shows that $\mathcal{J}_o$ is also a pdf, so we have $P_0(r,\ell) \geq 0$ and

$$2\pi\int_0^\infty rP_0(r,\ell)dr = 1. \qquad (2.7)$$

Before we examine the question of possible self-similarity in $\ell$, let us consider the shape of a prototypical axisymmetric pdf. It is described by its radial profile $P_0$ as illustrated schematically in Figure 3a. The profile has some generic properties. In particular, $P_0$ is never negative, and $P_0(r,\ell) \to 0$ as $r \to \infty$. The value at the origin, $P_0(0,\ell)$, is of interest. We have three possibilities: $P_0(0,\ell)$ could vanish, it might not exist, or it could be positive and finite. Figure 3a illustrates the last possibility. It is the simplest possibility and also the one selected by all our data. To describe the profile in Figure 3a, we need two parameters: size and steepness. A measure for the size is the radius $R_{half}$ of $\mathcal{I}_0$ at half the peak value. The steepness we can obtain from the derivative of $P_0$ at $R_{half}$. From the steepness we get the width $W$ of the transition region in $\mathcal{I}_0$; see Figure 3a. Both $R_{half}$ and $W$ may vary with $\ell$. If the ratio $W / R_{half}$ is independent of $\ell$ we have statistical self-similarity.

Next, let us consider how pdf's are usually rescaled. A semi-infinite axis, e.g., $r \geq 0$, needs a unit. Likewise, a full axis with given origin needs a unit. Working with pdf 's, that unit is usually obtained from the second order moment. For example, in the literature $\left\langle \left( \delta v_\parallel \right)^2 \right\rangle^{1/2}$ is used to normalize the pdf for $\delta v_\parallel$. The random variable $\delta v_\parallel$ is designed so that $\left\langle \delta v_\parallel \right\rangle \equiv 0$. Thus, even though that random variable can take any real value, the origin on the axis is already given. In general, however, the origin is not given on the full axis, but must be chosen. Thus, for the full axis we need to select both an origin $\mu$ and a unit $\sigma > 0$. Mean and standard deviation are the standard choices in statistics. If we change origin and unit on an axis, we perform a coordinate transformation called an affinity. Invariance after an appropriate affinity is affine symmetry. It is common in statistics. For example, affine symmetry occurs when multiple pdf's collapse onto a single graph after being plotted using mean and standard deviation for respectively centering and scaling. In the special case where the mean vanishes, the affine symmetry reduces to a simple rescaling and the symmetry is just "statistical self-similarity" (i.e., K41-scaling).

The obvious question is now whether we can expect affine symmetry in <u>our</u> turbulence statistics. If we have such symmetry, then we shall say that the turbulence has <u>*normal scaling*</u>. On the semi-infinite $r$-axis ($r \geq 0$), we need $\mu = 0$ for the affinity to map the semi-infinite axis onto itself. As discussed in connection with Figure 3a, the description of a monopole calls for two parameters, not just one. To have both $\mu$ and $\sigma$ free, we need a full axis. That is, we must map $r \geq 0$ onto a full axis, say the $\Omega -$ axis, with a non-linear transform. A logarithmic axis comes to mind, i.e.,

$$\Omega : r \rightarrow \Omega(r) = \ln r.  \tag{2.9}$$

In Appendix I, we show that there are no other meaningful choices for $\Omega(r)$ besides logarithmic functions. Figure 3b illustrates our prototypical monopole profile on a logarithmic axis. We notice immediately that mean and variance do not exist after the transformation. However, it is also clear that obvious choices exist for both $\mu$ and $\sigma$. For example, we could think of the radius $R_{half}$ as mapped to $\mu$ and $W$ to $\sigma$. There are no constraints on $\mu$ and $\sigma$, other than $\sigma > 0$. The place to look for affine symmetry is thus on the $\Omega -$ axis. The motivation for looking for symmetry is the old idea that the physics should be independent of scale in the inertial range. That was the idea behind K41. We are trying that idea again with our new statistical variables together with the Fourier notion of scale. Because our statistical variables are readily accessible in both shell model and DNS data, we can formulate a symmetry hypothesis and test it. Later we analyze its theoretical implications. The hypothesis reads:

**H1**: $P_0(r,\ell)$ scales normally in $\Omega = \ln r$.

Equivalently,

$$P_0(r,\ell) = C(\ell) f\left(\frac{\ln r - \mu(\ell)}{\sigma(\ell)}\right) \tag{2.10}$$

where $f$ is a similarity profile, while $\mu$ and $\sigma$ describe the affinity. It is not essential that $\mu$ and $\sigma$ are defined as in Figure 3b. (2.7) determines $C(\ell)$.

H1 is the key. Let us describe how H1 fares with computational data. We describe both shell model data and DNS data. H1 first emerged as an observation in high Reynolds number shell model data. The specific shell model we used was Zimin's model (Melander & Fabijonas 2001, 2003, Hussain and Zimin 1995). This model has two real variables similar to $\chi^+$ and $\chi^-$ per shell (i.e., per scale). The shell model was forced steadily at the largest scale and run into equilibrium 250,000 times to generate a solution ensemble at $R_\lambda \simeq 3 \times 10^6$. The statistical analysis of that ensemble follows exactly the framework laid out above and allows $P_o(r;\ell)$ to be constructed. The appropriate technique for doing that is to sample the random variable $\ln\left(X^2+Y^2\right)$ and construct its pdf through kernel density estimation; see Silverman (1990). From that pdf our radial profile $P_o(r,\ell)$ can be calculated. The result is Figure 4a, where $P_o(r,\ell)$ is plotted on a double logarithmic chart for various $\ell$. On the logarithmic chart (2.10) implies that the graphs collapse to one when the abscissa is stretched appropriately (i.e., by $\sigma(\ell)$) and the graphs then moved rigidly to the same location (i.e., shifts by $\mu(\ell)$ and $\ln C(\ell)$). The collapse is shown on the top in magnified form. For DNS data ($R_\lambda \approx 200$) we proceed the same way with $\ln\left(\left(\chi^+\right)^2+\left(\chi^-\right)^2\right)$ sampled from spherical shells $k_o-1/2 < \left|\vec{k}\right| \leq k_o+1/2$ in wave number space. Figure 4b shows the result for a $512^3$ dataset obtained from LANL. Importantly, the data collapses in the same way as for the shell model (although to a different curve). Hence, H1 is backed by both shell model and DNS data.

Finally, we consider the axisymmetric structure functions for $\mathcal{J}$. Because the area element in polar coordinates is $r\,dr\,d\theta$ we define the structure function of order $p$ as the $(p+1)th$ moment of $P_0(r,\ell)$:

$$S_p(\ell) \equiv \iint_{\mathbb{R}^2} \left(x^2+y^2\right)^{p/2} \mathcal{J}(x,y;\ell)\,dx\,dy = 2\pi\int_0^\infty r^{p+1}P_0(r,\ell)\,dr . \qquad (2.11)$$

With this definition we can state our second assumption.

**H2:** All convergent structure functions $S_p(\ell)$ are power laws in $\ell$ within the inertial range.

The assumption of power laws is classical in the literature on inertial range statistics and is backed by a substantial body of evidence of both experimental (e.g., Benzi et al 1995), computational (e.g., Boratav & Pelz 1997), and theoretical nature (Lundgren 2003). The literature, however, deals with the classical variables ($\delta_\parallel$ and $\delta_\perp$), not our new variables ($\chi^+$ and $\chi^-$). Thus, we supply computational data to substantiate H2 for our variables. The shell model data has a wide inertial range where power laws are evident; see Figure 5a. For the DNS data, the inertial range is much shorter because the Reynolds number is much lower. The standard technique is then ESS (see Benzi, et al 1995) which includes data from the dissipation range. ESS plots $S_p(\ell)$ against $S_3(\ell)$, instead of $\ell$, on a double logarithmic chart. The resulting graphs should then be straight lines. They are; see Figure 5b.

H2 allows us to write

$$S_p(\ell) = C_p \left( \ell/\ell_o \right)^{\zeta_p},\tag{2.12}$$

where $\zeta_p$ is the scaling exponent, $C_p$ the scaling coefficient, and $\ell_o$ is a reference length. We could absorb the factor $\ell_o^{-\zeta_p}$ into the constant $C_p$ without changing $S_p$. This fact merely shows that there is interdependence between the reference length and scaling coefficient. Once we specify $\ell_o$ the power laws (2.12) uniquely determine $C_p$. We will carry $\ell_o$ in our calculations as arbitrary until a natural length scale emerges.

With our data supporting H1 and H2, we study the Navier Stokes turbulence in lieu of these two assumptions. This is our link to the Navier Stokes equations. The interesting question is this: How restrictive is the combination of H1 and H2? The simple answer is: very restrictive; $P_o(r; \ell)$ must belong to a very small class of functions. The next three sections comprise a mathematical analysis to that effect. We find formulas for the structure functions and the underlying pdf. These formulas are all listed in the beginning of Section 6.

## 3. Deriving of the functional equation

The appropriate tool for analyzing the consequences of H1 and H2 is the Mellin transform:

$$\Phi(z) = \mathcal{M}\left[\phi(r); z\right] \equiv \int_0^\infty r^{z-1}\phi(r)dr \quad . \tag{3.1}$$

This transform deals with moments of a function defined on the positive real axis. If known, $\Phi(z)$ provides all moments of $\phi(r)$ – integer as well as fractional orders. Like the inverse Laplace transform, the inverse Mellin transform is a contour integral in the complex plane:

$$\phi(r) = \mathcal{M}^{-1}\left[\Phi(z); r\right] = \frac{1}{2\pi i}\int_{c-i\infty}^{c+i\infty} \Phi(z)r^{-z}dz \, . \tag{3.2}$$

The Mellin transform is related to both the Fourier and Laplace transforms; see Sneddon (1995) p. 41. In fact, operational rules for those transforms have Mellin counterparts. In particular, we need the rule:

$$\mathcal{M}\left[\phi\left((x/a)^{1/q}\right);z\right] = a^z q\Phi(qz)\,,\ a,q > 0 \tag{3.3}$$

For further details about the transform we refer to Oberhettinger (1974) or Paris & Kaminski (2001). The Mellin transform allows us to write the moments (2.11) in the form

$$S_p(\ell) \equiv 2\pi \int_0^\infty r^{p+1} P_0(r,\ell)\,dr = 2\pi \mathcal{M}\left[P_0(r,\ell); p+2\right]. \tag{3.4}$$

We introduce the composite function

$$g \equiv f \circ \ln \tag{3.5}$$

so that (2.10) becomes

$$P_0(r,\ell) = Cf\left(\frac{\ln r - \mu}{\sigma}\right) = Cf\left(\ln\left(re^{-\mu}\right)^{1/\sigma}\right) = C(\ell)g\left(\left(re^{-\mu(\ell)}\right)^{1/\sigma(\ell)}\right). \tag{3.6}$$

Defining $G$ as the Mellin transform of $g$

$$G(z) \equiv \mathcal{M}\left[g(r); z\right], \tag{3.7}$$

we use (3.4) and (3.3) to obtain

$$S_p(\ell) = 2\pi C\,\mathcal{M}\left[g\left(\left(re^{-\mu}\right)^{1/\sigma}\right); p+2\right] = 2\pi C e^{(p+2)\mu}\sigma G\left((p+2)\sigma\right). \tag{3.8}$$

In particular, we note that the identity

$$1 = S_0(\ell) = 2\pi C\,\mathcal{M}\left[P_0(r,\ell); 2\right] = 2\pi C e^{2\mu}\sigma G(2\sigma) \tag{3.9}$$

provides an expression for $C$, namely $C = 1/(2\pi\sigma e^{2\mu}G(2\sigma))$. Thus, we can eliminate $C$ from (3.8), whereby we obtain

$$S_p(\ell) = e^{p\mu}\frac{G\left((p+2)\sigma\right)}{G(2\sigma)}. \tag{3.10}$$

The next step is to eliminate $\mu$ from (3.10). For that purpose, we put $p = 3$ in (3.10) and use (2.12) to obtain

$$S_3(\ell) = C_3 \left( \frac{\ell}{\ell_o} \right)^{\zeta_3} = e^{3\mu} \frac{G(5\sigma)}{G(2\sigma)} \ , \tag{3.11}$$

and consequently

$$e^\mu = \left( C_3 \left( \frac{\ell}{\ell_o} \right)^{\zeta_3} \frac{G(2\sigma)}{G(5\sigma)} \right)^{1/3} . \tag{3.12}$$

Substituting (3.12) into (3.10), we find

$$S_p(\ell) = C_3^{\ p/3} \left( \frac{\ell}{\ell_o} \right)^{p\zeta_3/3} \left( \frac{G(2\sigma)}{G(5\sigma)} \right)^{p/3} \frac{G((p+2)\sigma)}{G(2\sigma)} . \tag{3.13}$$

(Note, we need no assumptions about the values of neither $\zeta_3$ nor $C_3$, and we could have used a different value of $p$ besides $p = 3$ to eliminate $\mu$ from 3.10.) Combining (2.12) and (3.13) yields an equation for the deviation from K41-scaling:

$$\left( \frac{G(2\sigma)}{G(5\sigma)} \right)^{p/3} \frac{G((p+2)\sigma)}{G(2\sigma)} = \frac{C_p}{C_3^{\ p/3}} \left( \frac{\ell}{\ell_o} \right)^{\zeta_p - p\zeta_3/3} . \tag{3.14}$$

Here, the left hand side depends on $\ell$ implicitly through $\sigma$. To simplify the right hand side of (3.14), we introduce the anomalous exponent

$$A(p) \equiv \zeta_p - p\zeta_3 / 3 \tag{3.15}$$

and the logarithm of the compensated coefficient

$$B(p) \equiv \ln \left( \frac{C_p}{C_3^{\ p/3}} \right) \tag{3.16}$$

The latter definition is permissible because $C_p$ is positive. This follows from (2.12) and the observation that $P_0(r; \ell)$ is not negative and therefore has positive moments (3.4). By construction,

$$A(0) = A(3) = B(0) = B(3) = 0 . \tag{3.17}$$

The next step will apply the logarithm to both sides of (3.14), so we define

$$\eta \equiv \ln \circ\, G \,. \tag{3.18}$$

This definition is permissible because $G > 0$ as can be seen from (3.8) by noting that $\sigma, C$ and $S_p(\ell)$ are all positive. After taken the logarithm, (3.14) translates to

$$\frac{p}{3}\big(\eta(2\sigma) - \eta(5\sigma)\big) + \eta\big((p+2)\sigma\big) - \eta\big(2\sigma\big) = A(p)(\ln \ell - \ln \ell_o) + B(p) \tag{3.19}$$

This equation is trivial if $\sigma(\ell)$ is constant. In that case, the left hand side is independent of $\ell$ and consequently $A(p) \equiv 0$ so that $\zeta_p = p\zeta_3/3$. Of course, that is K41 scaling which has no intermittency. Equation (3.19) gives no information in that case. Specifically, we do not obtain an expression for $B(p)$. Without $B(p)$, we do not get $C_p$. Without $C_p$, we do not have $S_p(\ell)$ which is needed to obtain $P(r,\ell)$ by the inverse Mellin transform of (3.7).

We can assume $\sigma(\ell)$ is not constant as otherwise there is no intermittency. Assuming smoothness, $\sigma \in C^1$, we can invert $\sigma(\ell)$ in intervals where it is monotone. In any such interval, we have:

$$w = \sigma(\ell) \quad \Leftrightarrow \quad \ell = \sigma^{-1}(w) \,. \tag{3.20}$$

By expressing (3.19) in terms of $w$, the left hand side becomes linear

$$\frac{p}{3}\big(\eta(2w) - \eta(5w)\big) + \eta\big((p+2)w\big) - \eta\big(2w\big) = A(p)(s(w) - \ln \ell_o) + B(p) \tag{3.21}$$

where

$$s(w) \equiv \ln(\sigma^{-1}(w)) \,. \tag{3.22}$$

Equation (3.21) is a functional equation for the unknown functions $\eta(w), s(w), A(p)$ and $B(p)$ in terms of the independent variables $p$ and $w$. The equation is the mathematical consequence of assuming nonlinear scaling exponents ($\zeta_p / p \neq Const$) in the presence of H1. The equation is our governing equation for intermittency. In the case of K41 scaling, which is equivalent to statistical self-similarity and no intermittence, the equation is trivially satisfied.

## 4. Solving the functional equation

Under the provision of sufficient smoothness ($\eta \in C^3; A, B \in C^2; s \in C^1$) we can find the complete solution to (3.21) in closed form. The smoothness conditions are inconsequential for our physical problem, but mathematically they exclude "pathological solutions." Our method for solving equation (3.21) consists of transforming it into one whose solution is known. This process requires us to introduce yet more functions and variables, but the effort pays off with the certainty of knowing the complete solution. In contrast, one can also obtain solutions by inspection (e.g., $\eta$ being a power or a logarithm solves 3.21), but then the question of completeness remains open. Our task in this section is thus to find $\eta(w), s(w), A(p)$ and $B(p)$ in closed form.

We start by differentiating (3.21) twice with respect to $p$:

$$w^2 \eta''\big((p+2)w\big) = A''(p)\big(s(w) - \ln \ell_o\big) + B''(p). \tag{4.1}$$

Differentiating again with respect to $w$ yields

$$\frac{\partial}{\partial w}\left(w^2\eta''\left((p+2)w\right)\right) = \frac{\partial}{(p+2)^2\partial w}\left((p+2)^2 w^2\eta''\left((p+2)w\right)\right) = A''(p)s'(w).\qquad(4.2)$$

To simplify the left hand side, we define a new function:

$$L(\chi) \equiv \left(\chi^2\eta''(\chi)\right)'.\qquad(4.3)$$

Upon multiplication by $p+2$, (4.2) takes the form

$$L\left((p+2)w\right) = (p+2)A''(p)s'(w).\qquad(4.4)$$

To simplify the right hand side of (4.4), we define two new functions:

$$R_1(\chi) \equiv \chi A''(\chi-2), \qquad R_2(\chi) \equiv s'(\chi).\qquad(4.5)$$

Thereby, (4.4) takes the simple form

$$L(xy) = R_1(x)R_2(y), \qquad x, y > 0.\qquad(4.6)$$

The solution to this functional equation consists of those functions $L, R_1$ and $R_2$ for which (4.6) is satisfied for all positive $x$ and $y$. As the reader may imagine, (4.6) can be found in the literature on functional equations. In fact, (4.6) is one of Pexider's equations which in turn is a generalization of Cauchy's equation $\phi(x+y) = \phi(x) + \phi(y)$; see Aczel (1966) p.141 and p.37.

Using our notation, Theorem 4 page 144 in Aczel (1966) states: "The general solutions, with $L$ continuous at a point, of (4.6) are

$$L(t) = abt^{\beta-1}, \ R_1(t) = at^{\beta-1}, \ R_2(t) = bt^{\beta-1} \ (t>0)\qquad(4.7)$$

supplemented by the trivial solutions." Here $a$, $b$, $\beta$ are arbitrary constants. In (4.7), we write the exponent as $\beta-1$, rather than $\beta$, so as to streamline the subsequent analysis. Evidently, we can transfer a constant factor from $R_1$ to $R_2$ without changing $L$. Thus, a redundancy exists between constants $a$ and $b$ (i.e., only their product enters into $L$.) We will remove this redundancy when we develop expressions for $\eta(w), s(w), A(p)$ and $B(p)$.

Let us deal with the trivial solutions to (4.6) first. They are respectively ($L(t) \equiv R_1(t) \equiv 0$, $R_2(t)$ arbitrary) and ($L(t) \equiv R_2(t) \equiv 0$, $R_1(t)$ arbitrary). For the first solution, $R_1(t) \equiv 0$ so that (4.5) yields $A''(p) \equiv 0$ which because of (3.17) implies $A(p) \equiv 0$ or, equivalently, $\zeta_p = p\zeta_3/3$ via (3.15). For the second solution, $R_2(t) \equiv 0$. That also means $\zeta_p = p\zeta_3/3$ because (4.5), (3.22) and (3.20) yield $\sigma = const$. Thus, the trivial solutions all correspond to K41-scaling for which our analysis yields nothing. The trivial solutions to (4.6) are therefore irrelevant.

The next task is to translate the solution (4.7) into expressions for $A(p), B(p), s(w)$ and $\eta(w)$. Combining (4.7) with (4.3) and (4.5) we obtain three differential equations:

$$\left(\chi^2 \eta''(\chi)\right)' = ab\chi^{\beta-1}, \tag{4.8}$$

$$A''(\chi-2) = a\chi^{\beta-2}, \tag{4.9}$$

$$s'(\chi) = b\chi^{\beta-1}. \tag{4.10}$$

In solving these equations, two special cases, $\beta = 0, 1$, arise from the fact that $\int x^m dx$ is a power when $m \neq -1$, but a logarithm when $m = -1$. Here, we focus on the main case $\beta \neq 0, 1$. Appendix II deals with the two special cases. Since we now know $L(\chi)$, we can backtrack from (4.3) to (4.1). Thus, we first integrate (4.8) once to obtain

$$\chi^2 \eta''(\chi) = \frac{ab}{\beta} \chi^\beta + c_1 \qquad (4.11)$$

where $c_1$ is an integration constant. Then, we substitute $\chi = (p+2)w$ into (4.11) and divide by $(p+2)^2$. The result is

$$w^2 \eta''((p+2)w) = \frac{ab}{\beta}(p+2)^{\beta-2} w^\beta + \frac{c_1}{(p+2)^2}. \qquad (4.12)$$

Comparing (4.1) and (4.12), we see that the left hand sides are equal. Thus,

$$A''(p)(s(w) - \ln \ell_o) + B''(p) = \frac{ab}{\beta}(p+2)^{\beta-2} w^\beta + \frac{c_1}{(p+2)^2}. \qquad (4.13)$$

From (4.9) we have

$$A''(p) = a(p+2)^{\beta-2} \qquad . \qquad (4.14)$$

Substituting (4.14) into (4.13) and rearranging, we find

$$a(p+2)^{\beta-2}\left(s(w) - \ln \ell_o - \frac{b}{\beta} w^\beta\right) = \frac{c_1}{(p+2)^2} - B''(p). \qquad (4.15)$$

The right hand side of (4.15) is independent of $w$. Consequently, (4.15) separates into two equations by means of a separation constant $s_0$:

$$s(w) - \ln \ell_o - \frac{b}{\beta} w^\beta = s_o, \qquad (4.16)$$

$$B''(p) = \frac{c_1}{(p+2)^2} - a(p+2)^{\beta-2} s_o. \qquad (4.17)$$

It is important to recognize the significance of $s_0$. This constant clearly affects the solution $B(p)$ we obtain by solving the differential equation (4.17) with auxiliary conditions provided by (3.17). In turn, the constant $s_0$ affects the expression for the scaling coefficient $C_p$; see (3.16). Moreover, (4.16) shows that $s_0$ influences the expression for $s(w) \equiv \ln \ell$ and hence $w \equiv \sigma(\ell)$, i.e., (3.20) and (3.22), but the influence is only the through the combined constant $s_0 + \ln \ell_o$. We can trace the dependence on $s_0$ all the way back to the expression for the moments $S_p(\ell)$. By doing so, we find that $S_p(\ell)$ is independent of $s_0$. Moreover, we recognize that $s_0$ exists because $C_p$ depends on the reference scale for $\ell$; see the discussion below (2.12). Previously, we introduced that reference scale as $\ell_o$, but left it otherwise unspecified. Now, we have arrived at the point where we can choose $\ell_o$ to be an intrinsic length scale for the inertial range scaling regime. This choice of $\ell_o$ emerges by choosing $s_0 = 0$ so that the right hand side in (4.17) contain no component of (4.14). This choice results in the simplest expression for $C_p$, i.e., one that is not polluted by a factor $(const)^{\zeta_p}$. We shall also see below that $\ell_o$, thus chosen, marks the infrared termination of the scaling regime.

Solving (4.14) and (4.15) with auxiliary conditions (3.17) yields respectively

$$A(p) = \frac{a}{3(\beta-1)\beta}\left(3(p+2)^{\beta} + p(2^{\beta} - 5^{\beta}) - 3 \times 2^{\beta}\right) , \quad \beta \neq 0,1 \qquad (4.18)$$

and

$$B(p) = c_1\left(-\ln(p+2) + p\ln\left((5/2)^{1/3}\right) + \ln 2\right) \quad \beta \neq 0,1 . \qquad (4.19)$$

Combining (3.20) and (3.22), we have $s(w) = \ln \ell$ and $w = \sigma(\ell)$ so that (4.16) reads

$$\sigma^{\beta} = \frac{\beta}{b}\left(\ln \ell - \ln \ell_o\right) , \quad \beta \neq 0 . \qquad (4.20)$$

The right hand side must be positive to assure a positive single valued solution $\sigma$. In studying small scales, $\ell < \ell_o$, we must therefore require $\beta / b < 0$. (In studying an infrared cascade one would make the opposite choice. This would be the situation if one were to study scales larger than the forcing scale - like in the cascade from small to large scales advocated by C. Gibson (1996).) We remove the redundancy between $a$ and $b$ in (4.7) by taking

$$b = -\beta.$$ 
(4.21)

It is the same redundancy that exists between $f$ and $\sigma$ in (2.10). With that redundancy removed, we have

$$\sigma = \left( \ln \ell_o - \ln \ell \right)^{1/\beta}.$$ 
(4.22)

Here, $\ln \ell_o$ marks the virtual origin for the inertial range scaling laws. The scaling laws can not be continued to scales larger than $\ell_o$. Thus, our $\ell_o$ is truly an intrinsic length scale for the inertial range unlike the integral scale used in K41.

Substituting (4.21) into (4.11), we integrate twice to obtain:

$$\eta \left( \chi \right) = -\frac{a \chi^{\beta}}{\left( \beta - 1 \right) \beta} - c_1 \ln \chi + c_2 \chi + c_3, \quad \beta \neq 0, 1.$$ 
(4.23)

Here $c_2$ and $c_3$ are arbitrary integration constants. In fact, $c_2 \chi + c_3$ is the homogenous solution to (4.11) as well as to (3.21). Because (3.21) is a linear equation in $\eta$, a homogeneous solution $\left( c_2 \chi + c_3 \right)$ can be added to any particular solution so as to produce another solution to (3.21). We note that $S_p \left( \ell \right)$ can be constructed from the functions on the right hand side of (3.21) using (2.12) together with (3.15), (3.16) and the fact that $\ln \ell = s(w)$. Since $c_2 \chi + c_3$ does not affect the right hand side of (3.21), it then follows that $c_2 \chi + c_3$ leaves $S_p(\ell)$ invariant. Serving as a "normalization" or "gauge" of $\eta$, the homogenous solution $c_2 \chi + c_3$ changes both $\mu$ and $G$ within (3.10) in such a way as to leave $S_p(\ell)$ invariant. Thus, we are free to choose $c_2$ and $c_3$ so that we obtain the nicest expressions for $\mu$ and $G$. Our choice is

$$c_2 = c_3 = 0. \tag{4.24}$$

Specifically, $e^\mu$ is a power law when $c_2 = 0$. That is, combining (4.24) with (3.12), (3.18), (4.20) and (4.23) yields

$$e^\mu = \left( C_3 \left( \frac{\ell}{\ell_o} \right)^{\zeta_3} \exp\left( \eta(2\sigma) - \eta(5\sigma) \right) \right)^{1/3} = \left( C_3 5^{c_1} / 2^{c_1} \right)^{1/3} \left( \frac{\ell}{\ell_o} \right)^{(\zeta_3 - \gamma)/3} \tag{4.25}$$

where $\gamma = \dfrac{a\left( 2^\beta - 5^\beta \right)}{(\beta - 1)\beta}$.

# 5. Properties of the solution

Section 4 gave us the solution to the functional equation (3.21). Let us now find the similarity profile $g$ in (3.6). Using (3.18), (4.23) and (4.24), we have

$$G(z) = \exp(\eta(z)) = z^{-c_1} e^{\kappa z^\beta}, \quad \kappa \equiv -\frac{a}{\beta(\beta - 1)}, \quad \beta \neq 0, 1, \tag{5.1}$$

so that inversion of (3.7) yields

$$g(r) = \mathcal{M}^{-1}\left[ G(z); r \right] = \mathcal{M}^{-1}\left[ z^{-c_1} e^{\kappa z^\beta}; r \right], \quad \beta \neq 0, 1. \tag{5.2}$$

In this section, we analyze the two expressions (5.1) and (5.2). (Appendix II contains the corresponding analysis for the special cases $\beta = 0, 1$.) There are several important issues with which we must deal. Starting with the inversion process, the inverse Mellin transform must exist and produce a real function (i.e., zero imaginary part). That is, the integral (3.2) must converge to a real value for all $r \geq 0$. That imposes a selection criterion on the parameters $a, \beta, c_{-1}$. Except for a few exceptional cases $\beta = 1/3, 1/2,$ and $2$, the literature does not provide the inverse transform (5.2). Consequently, we must introduce a new class of special functions. To analyze asymptotic behavior and construct numerical algorithms for these functions, we reduce them to a normal form free of removable constants. Another important issue is the value of the constant $c_1$, we shall see that it determines the behavior of $g(r)$ in the limit of small $r$. As is evident from (3.16) and (4.19), $c_1$ affects the scaling coefficient $C_p$, but the scaling exponent $\zeta_p$ is independent of $c_1$; see (3.15) and (4.18). Analysis of the other azimuthal modes in (2.5) leads to same functional equation (3.21) and therefore to the same formulas for $\zeta_p$ and $C_p$. However, the value of $c_1$ will differ between various azimuthal modes in (2.5). Since $c_1$ does not enter in the expression for the scaling exponent $\zeta_p$, all azimuthal modes can have common scaling exponents. That allows the joint pdf $J$ to have scaling laws. Here, we show that $c_1 = 1$ for the axisymmetric mode in (2.5). The final issue is whether $J_o$, as constructed from $g(r)$ via (2.6) and (3.6), is actually a pdf. Because $S_o(\ell) = 1$ is already arranged by (3.9), $J_o$ is a pdf if and only if $g(r) \geq 0$. This condition further restricts the parameter $\beta$.

## 5.1 Existence of the inverse Mellin transform and its normal form.

First, let us ensure that we obtain a real, rather than a complex, function $g(r)$ from (5.2).

Using the polar notation $z = \rho e^{i\theta}$, (5.1) reads

$$G(z) = z^{-c_1} e^{\kappa z^\beta} = \rho^{-c_1} e^{\kappa \rho^\beta \cos(\beta\theta)} e^{-ic_1\theta + i\kappa\rho^\beta \sin(\beta\theta)}. \tag{5.3}$$

$z^\beta$ is multi-valued except at $z = 0$. Thus, we have to introduce a branch cut emerging from $z = 0$. We place this cut along the negative real axis and select $\theta$ as the principal argument of $z$:

$$-\pi < \theta < \pi. \tag{5.4}$$

Then it follows from (5.3) that $G(\overline{z}) = \overline{G(z)}$. We also have $r^{-\overline{z}} = \overline{r^{-z}}$. Thus, if the integral (3.2) converges, it is real, namely:

$$\frac{1}{2\pi i} \int_{c-i\infty}^{c+i\infty} G(z) r^{-z} dz = \frac{1}{\pi} \int_0^\infty \text{Re}(G(c+iy) r^{-c-iy}) dy \tag{5.5}$$

where $c > 0$ so as to avoid intercepting the branch cut. All other branches of $z^\beta$ make $g(r)$ complex.

Convergence is our next topic. Along the line $z = c + iy$, we have

$$\left| G(z) r^{-z} \right| = \rho^{-c_1} e^{\kappa \rho^\beta \cos(\beta\theta)} r^{-c}. \tag{5.6}$$

With $\beta \neq 0,1$, the exponential function $e^{\kappa \rho^\beta \cos(\beta\theta)}$ determines whether the integral (5.5) converges or diverges. Along $z = c + iy$, we have $\theta \to \pm\pi/2$ as $z \to \infty$. Thus, (5.5) converges precisely when

$$\kappa \cos\left(\beta\pi/2\right) = \frac{-a}{(\beta-1)\beta} \cos\left(\beta\pi/2\right) < 0. \tag{5.7}$$

We can establish that $a$ is never positive. Using (3.15) and (4.14), we have $\zeta_p'' = A''(p) = a(p+2)^{\beta-2}$. Moreover, from the analysis below (1.10) we know $\zeta_p'' \leq 0$ (on the ultraviolet side). Consequently,

$$a < 0. \tag{5.8}$$

The combination of (5.7) and (5.8) shows that $\left(\cos\left(\beta\pi/2\right)\right)/\left(\left(\beta-1\right)\beta\right)$ must be negative. That eliminates the intervals $..., -9 < \beta < -7, \quad -5 < \beta < -3, \quad -1 < \beta < 0, \quad 3 < \beta < 5, \; 5 < \beta < 7, ...$.

Finally, let us express $G(z)$ in normal form so that we can avoid numerical computation of inverse Mellin transforms otherwise obtainable through scaling. Specifically, we can express $G(z)$ in terms of another function (the normal form) that does not involve the constant $\kappa$. Using (5.8), we have $\kappa = \dfrac{|a|}{\beta\left(\beta-1\right)} = |\kappa|\,\mathrm{sgn}\left(\beta(\beta-1)\right)$ so that we can write

$$G(z) = z^{-c_1} e^{|\kappa| z^\beta \,\mathrm{sgn}(\beta(\beta-1))} = |\kappa|^{c_1/\beta}\left(|\kappa|^{1/\beta} z\right)^{-c_1} e^{\left(|\kappa|^{1/\beta} z\right)^\beta \,\mathrm{sgn}(\beta(\beta-1))}. \tag{5.9}$$

This observation allows us to express the inverse Mellin transform in normal form. Defining

$$\mathcal{B}\left(z\right) \equiv z^{-c_1} e^{z^\beta \,\mathrm{sgn}(\beta(\beta-1))} \tag{5.10}$$

and

$$b(r) \equiv \mathcal{M}^{-1}\left[\mathcal{B}(z); r\right], \tag{5.11}$$

equation (5.9) becomes

$$G(z) = z^{-c_1} e^{|\kappa| z^\beta \,\mathrm{sgn}(\beta(\beta-1))} = |\kappa|^{c_1/\beta}\, \mathcal{B}\left(|\kappa|^{1/\beta} z\right). \tag{5.12}$$

Moreover, using (3.3) we obtain

$$g(r) = \left|\kappa\right|^{c_1/\beta} \mathcal{M}^{-1}\left[\mathcal{B}\left(\left|\kappa\right|^{1/\beta} z\right); r\right] = \left|\kappa\right|^{(c_1-1)/\beta} \, b(r^{\left|\kappa\right|^{-1/\beta}}) \,. \tag{5.13}$$

With appropriate restrictions imposed on $\beta$ and $c_1$, (5.10) and (5.11) form a new class of special functions. Using them, we recover $g(r)$ and the pdf for the inertial range.

## 5.2 Determining the value of $c_1$.

Consider the azimuthal expansion (2.5) in the limit of small $r$. If $J$ has to be smooth at $r = 0$ then $P_1, P_2, \ldots$ must all vanish there. The value of $J$ at the origin is therefore determined solely by the axisymmetric term $P_0$. It has a finite positive value precisely when $g(0)$, and hence $b(0)$, does; see (3.6) and (5.13) respectively. Thus, we require that $b(r)$ has a limit as $r \to 0^+$. Moreover, that limit can be neither zero nor infinity; see Figure 3a.

To facilitate the analysis for small $r$, we introduce the function

$$\tilde{b}(r,\delta) \equiv \mathcal{M}^{-1}\left[\frac{z+\delta}{z}\mathcal{B}(z+\delta); r\right], \quad \delta > 0 \,. \tag{5.14}$$

This function has a finite value at $r = 0$, as we shall see below. Moreover, as $\delta \to 0^+$ for $r > 0$, we have $\tilde{b}(r,\delta) \to b(r)$, but not necessarily uniformly in $r$. To find $\tilde{b}(0,\delta)$ we use the convolution theorem for Mellin transforms (Oberhettinger 1974),

$$\mathcal{M}^{-1}\left[\Phi_1(z)\Phi_2(z); r\right] = \int_0^\infty t^{-1}\phi_1\left(r/t\right)\phi_2\left(t\right) dt \,, \tag{5.15}$$

with $\Phi_1(z) = (z+\delta)\mathcal{B}(z+\delta)$ and $\Phi_2(z) = 1/z$. We denote the inverse functions by $\phi_1$ and $\phi_2$ respectively. For the latter, we have

$$\phi_2(r) = \mathcal{M}^{-1}[1/z; r] = \begin{cases} 1, r < 1 \\ 0, r > 0 \end{cases} \quad ; \tag{5.16}$$

see Oberhettinger (1974).

Using (5.14) and (5.15), we can write

$$\tilde{b}(r,\delta) = \int_0^\infty t^{-1}\phi_1(r/t)\phi_2(t)\,dt = \int_0^1 t^{-1}\phi_1(r/t)\,dt = \int_r^\infty u^{-1}\phi_1(u)\,du. \tag{5.17}$$

Consequently,

$$\tilde{b}(0,\delta) = \int_0^\infty u^{-1}\phi_1(u)\,du = \mathcal{M}[\phi_1(z); 0] = \Phi_1(0) = \delta\mathcal{B}(\delta) = \delta^{1-c_1}e^{\delta^\beta\,\mathrm{sgn}(\beta(\beta-1))}, \tag{5.18}$$

on account of (3.1), the definition of $\Phi_1(z)$, and (5.10). We can now examine the behavior of $\tilde{b}(r,\delta)$ in the limit $\delta \to 0+$. There are two cases to consider: $\beta < 0$ and $\beta > 0$. In the first case, $\tilde{b}(0,\delta) = \delta^{1-c_1}e^{\delta^\beta\,\mathrm{sgn}(\beta(\beta-1))} \to \infty$ as $\delta \to 0^+$ regardless the value of $c_1$. Thus, we rule negative $\beta$ impermissible. In the second case, we have:

$$\tilde{b}(0,\delta) = \delta^{1-c_1}e^{\delta^\beta\,\mathrm{sgn}(\beta(\beta-1))} \to \begin{cases} 0, c_1 < 1 \\ 1, c_1 = 1 \\ \infty, c_1 > 1 \end{cases} \text{ as } \delta \to 0^+. \tag{5.19}$$

Only $c_1 = 1$ provides us with a finite positive limit. Since $\tilde{b}(r,\delta) \to b(r)$ for $r > 0$ and $b(r)$ is continuous it follows that only $c_1 = 1$ gives a finite nonzero value of $b(0)$.

Moreover, that value is $b(0) = 1$. The results of the analysis in Section 5.2, can be summarized as

$$c_1 = 1, \quad \beta > 0, \quad b(0) = 1. \tag{5.20}$$

## 5.3 Non-negative b(r).

Our last requirement is that $g(r)$ is non-negative so that $J_0$ is a pdf. On account of (5.13), $g(r)$ and $b(r)$ have the same sign. Using the result (5.20), $b(r)$ depends only on one parameter: $\beta$; see (5.10) and (5.11). Because of the requirement $b(r) \geq 0$, we must reject those values of $\beta$ for which $b(r)$ takes negative values. Below we analyze the sign of $b(r)$. The analysis splits into many separate cases. Before looking into these technicalities, let summarize the results of the analysis. We show analytically that $b(r) \geq 0$ when $0 < \beta \leq 2$. We also demonstrate analytically that $b(r) \geq 0$ when $2 < \beta < 3$ and $r \geq 1$, but we must rely on numerical techniques for $r < 1$. For larger $\beta$, $b(r)$ either does not exist (i.e., for $3 \leq \beta \leq 5, 7 \leq \beta \leq 9$, …) or numerical calculations show $b(r)$ to be an oscillatory function with both positive and negative values (i.e., for $5 < \beta < 7, 9 < \beta < 11$, …). Moreover, Appendix II rules out $\beta = 0$, but allows $\beta = 1$. Thus, we have $b(r)$ non-negative precisely when $0 < \beta < 3$. The corresponding graphs are shown Figure 6.

Our analysis uses the formula

$$b(r) = \frac{1}{2\pi i} \int_{c-i\infty}^{c+i\infty} \mathcal{B}(z) r^{-z} dz = \frac{1}{2\pi i} \int_{c-i\infty}^{c+i\infty} z^{-1} e^{z^\beta \operatorname{sgn}(\beta-1) - z \ln r} dz, \quad c > 0. \tag{5.21}$$

Because $z^\beta = \rho^\beta e^{i\beta\theta}$ is an oscillatory function of $\theta$, the behavior of the integrand becomes increasingly more complicated as $\beta$ increases. Our analysis therefore splits into several separate cases according to the sign of $\ln r$ and the value of $\beta$.

### 5.3.1 Case $0 < \beta < 1$ and $r > 1$.

Using $z = \rho e^{i\theta}$, we have

$$\text{Re}(-z^\beta - z\ln r) = -\rho^\beta \cos\beta\theta - (\ln r)\rho\cos\theta \quad . \tag{5.22}$$

The last term dominates when $\rho$ is large. Thus, (5.22) tends to minus infinity as $\rho \to \infty$ when $\cos\theta > 0$. Consequently, the integrand in (5.21) decays faster than algebraically in the right half plane. With the only singularity located at the origin and the branch cut running along the negative real axis, the integrand is analytic in the right half plane. Thus, we can close the integration loop (5.21) with a large semicircle to the right and use Cauchy integral theorem to show that (5.21) vanishes identically. Thus, we have

$$b(r) \equiv 0, \quad 0 < \beta < 1 \text{ and } r > 1. \tag{5.23}$$

This means compact support for $b(r)$.

### 5.3.2 Case $0 < \beta < 1$ and $r < 1$.

With $\ln r$ negative, the integrand in (5.21) decays faster than algebraically only in the left half plane; see (5.22). So we can not close the integration loop with a right semicircle. Instead, we consider the curve defined by

$$\text{Im}(-z^\beta - z\ln r) = -\rho^\beta \sin\beta\theta - (\ln r)\rho\sin\theta = 0. \tag{5.24}$$

Along this curve the integrand in (5.21) does not oscillate. Rewriting (5.24) we obtain the steepest decent contour $\mathcal{S}$:

$$\mathcal{S}: \rho^{1-\beta} \ln(r^{-1}) = \begin{cases} \dfrac{\sin \beta \theta}{\sin \theta}, & \theta \neq 0 \\ \beta, & \theta = 0 \end{cases}, \tag{5.25}$$

The right hand side of (5.25) is positive for $-\pi < \theta < \pi$ and becomes unbounded precisely as $\theta \to \pm \pi$. Thus, $\mathcal{S}$ consists of a single arc that runs to infinity in both directions as illustrated in Figure 7a. Since $\mathcal{S}$ runs to infinity in the left half plane, we can deform the integration path in (5.21) to $\mathcal{S}$. It is thus possible to express $b(r)$ as an integral over $\theta$. For this purpose, we use the fact that (5.25) implies $\rho(\theta) = \rho(-\theta)$ so $\text{Re}(-z^\beta - z \ln r) = -\rho^\beta \cos \beta \theta - (\ln r) \rho \cos \theta$ is an even function of $\theta$. Moreover, $\rho'(\theta) = -\rho'(-\theta)$ so $\rho'(\theta)/\rho(\theta)$ is odd. Using $dz/z = (i + \rho'(\theta)/\rho(\theta))d\theta$, we thereby obtain

$$b(r) = \frac{1}{\pi} \int_0^\pi e^{-\rho^\beta \cos \beta \theta - (\ln r) \rho \cos \theta} d\theta, \qquad 0 < \beta < 1 \text{ and } r < 1. \tag{5.26}$$

Here the integrand is always positive, so $b(r) > 0$. Previously, see (5.20), we found $b(0) = 1$. Moreover, continuity at $r = 1$ yields $b(1) = 0$. (Both of these values can also be obtained directly from 5.21).

Since $b(r)$ does not take negative values, all $0 < \beta < 1$ are permissible. Corresponding graphs of $b(r)$ are shown in Figure 6a. In one particular case, $\beta = 1/2$, we have an expression in closed form Oberhettinger (1974):

$$\flat(r) = \begin{cases} \operatorname{erfc}\left(-1/\ln r\right), r < 1 \\ 0, r \geq 0 \end{cases}, \quad \beta = 1/2. \tag{5.27}$$

Here $\operatorname{erfc}$ is the complementary error function.

### 5.3.3 Case $1 < \beta < 3$ and $r > 1$.

With $1 < \beta$ the phase in (5.21) vanishes when $\operatorname{Im}(z^{\beta} - z \ln r) = 0$. Rearranging leads to the steepest decent contour:

$$S: \frac{\rho^{\beta-1}}{\ln(r)} = \begin{cases} \dfrac{\sin\theta}{\sin\beta\theta}, \theta \neq 0 \\ 1/\beta, \theta = 0 \end{cases}. \tag{5.28}$$

Both the right hand side and $\ln r$ are positive for $-\pi/\beta < \theta < \pi/\beta$. Thus, the curve $S$ consists of a single arc with asymptotes $\theta = \pm\pi/\beta$; see Figure 7b. Since $\operatorname{Re}(z^{\beta} - z \ln r) = \rho^{\beta}\cos\beta\theta - (\ln r)\rho\cos\theta$, we see that the integrand in (5.21) decays faster than algebraically when $\cos\beta\theta < 0$, but becomes unbounded when $\cos\beta\theta > 0$. Because $\cos\beta\theta < 0$ for all $\theta$ in the interval between $\pi/2$ and $\pi/\beta$, the integration path in (5.21) can be deformed to $S$; see Figure 7b. Using the fact that $\rho(\theta)$ is an even function we obtain

$$\flat(r) = \frac{1}{\pi} \int_0^{\pi/\beta} e^{\rho^{\beta}\cos\beta\theta - (\ln r)\rho\cos\theta} d\theta, \qquad 1 < \beta < 3 \text{ and } r > 1. \tag{5.29}$$

Here the integrand is positive. Consequently, $\flat(r)$ is also positive.

### 5.3.4 Case $1 < \beta < 2$ and $r < 1$.

Again we consider the vanishing phase, $\mathrm{Im}(z^\beta - z \ln r) = 0$, and again this equation can be rewritten as

$$\frac{\rho^{\beta-1}}{\ln(r)} = \frac{\sin\theta}{\sin\beta\theta}. \tag{5.30}$$

However, now $\ln r < 0$, so for $\rho^{\beta-1}$ to be positive the right hand side must be negative. That gives us two separate intervals $-\pi < \theta < -\pi/\beta$ and $\pi/\beta < \theta < \pi$. Correspondingly, we have two arcs in the complex plane. One arc has $\theta = \pi/\beta$ as an asymptote and runs from infinity to zero. The other also connects zero to infinity, but has $\theta = -\pi/\beta$ as its asymptote. To avoid the branch cut, we join the two arcs with a circular loop around zero to form a single curve $\mathcal{S}$; see Figure 7c. With increasing $\theta$, we traverse both arcs from top to bottom. We must reverse this orientation before we can deform the integration path in (5.21) to $\mathcal{S}$. The deformation is permissible because the integrand decays faster than algebraically for $-\pi/2 \leq \theta \leq -\pi/\beta$ and $\pi/\beta \leq \theta \leq \pi/2$; see Figure 7c. From the figure and the fact that that $\rho(\theta)$ is an even function, we find

$$b(r) = 1 - \frac{1}{\pi} \int_{\pi/\beta}^{\pi} e^{\rho^\beta \cos\beta\theta - (\ln r)\rho\cos\theta} d\theta, \qquad 1 < \beta < 2 \text{ and } r < 1. \tag{5.31}$$

To show that $b(r) \geq 0$, we establish that $b(1) \geq 0$ and $b'(r) < 0$. The first part follows from continuity at $r = 1$ and the fact $\lim_{r\to 1+} b(r) > 0$; see (5.29). One can also compute $b(1) = 1/\beta$ directly from (5.21). As for the second part, we need an expression for $b'(r)$. For this purpose, we use the operational rule (Oberhettinger 1974)

$$\mathcal{M}\left[r b'(r); z\right] = -z\mathcal{B}(z) = -e^{z^\beta}. \tag{5.32}$$

The contour integral for $-rb'(r) = \mathcal{M}^{-1}\left[e^{z^\beta}; z\right]$ has the same convergence properties as (5.21). Moreover, the zero phase contour remains unchanged, so we can use the same integration path $\mathcal{S}$; see Figure 7c. This time, the contribution from the circular loop around zero vanishes so that

$$rb'(r) = \frac{1}{\pi}\int_{\pi/\beta}^{\pi} e^{\rho^\beta \cos\beta\theta - (\ln r)\rho\cos\theta}\left(\rho'\sin\theta + \rho\cos\theta\right)d\theta, \qquad 1 < \beta < 2 \text{ and } r < 1. \quad (5.33)$$

In this integral, $\rho'(\theta)$ and $\cos\theta$ are both negative, while $\rho(\theta)$ and $\sin\theta$ are both positive; see Figure 7c. Consequently, the integrand in (5.33) is negative. It follows that $b'(r) < 0$. Combining this result with with $b(1) = 1/\beta > 0$ we obtain

$$b(r) > 0 \qquad 1 < \beta < 2 \text{ and } r < 1. \quad (5.34)$$

Incidentally, (5.33) has a finite negative limit as $r \to 0^+$. Thus, the profile $P_0(r;\ell)$ has a cups at the origin.

### 5.3.5 Case $\beta = 2$.

In this case the inverse Mellin transform is known in closed form (e.g. Oberhettinger 1974)

$$b(r) = \frac{1}{2}\text{erfc}\left(\frac{\ln r}{2}\right), \quad \beta = 2. \quad (5.35)$$

### 5.3.6 Case $2 < \beta < 3$ and $r < 1$.

As $\beta$ increases beyond two, the zero phase contour (5.29) separates into two disjoint pieces as shown in Figure 7d. Consequently, the argument used in Section 5.3.5 no longer works.  However, numerical calculations show $b(r) > 0$.  Graphs of $b(r)$ are shown in Figure 6d. Note the damped log-periodic oscillation.

**5.3.7** Case $5 < \beta$.

The zero phase contour now consists of multiple pieces regardless  of whether $r > 1$ or $r < 1$; see Figure 7e.  In particular, we note that for $r > 1$ the integration path in (5.21) can not be deformed to curve given by (5.28). Such a deformation would namely cross one of the regions where the integrand becomes unbounded. Numerical evaluation of (5.21) shows that  $b(r)$ is oscillatory with negative values.

# 6. Summary, Discussion and Conclusion

We have observed affine symmetry in computational data and formulated a corresponding   hypothesis (H1) for the inertial range.  Combining H1 with the traditional assumption (H2) of scaling laws for the structure functions, we obtain a potent analytical setup.  A very limited class of statistics satisfies both H1 and H2. We found this class by analytic means in Sections 3-5. Except for a few undetermined constants, we consequently know scaling exponents and coefficients for all structure functions of the random variable $R = \left( \left( \chi^- \right)^2 + \left( \chi^+ \right)^2 \right)^{1/2}$.  Correspondingly, we also know the underlying pdf.  Let us summarize the analysis by listing a complete set of formulas.

Our inertial range scaling laws read $\left\langle R^p \right\rangle = S_p \left( \ell \right) = C_p \left( \ell / \ell_o \right)^{\zeta_p}$, where for all  $p > -2$,

$$C_p = \frac{2}{p+2}\left(\frac{5C_3}{2}\right)^{p/3} \tag{6.1}$$

and

$$\zeta_p - \frac{p\zeta_3}{3} = \frac{a}{3}\begin{cases} \dfrac{3(p+2)^{\beta} + p(2^{\beta}-5^{\beta}) - 3\times 2^{\beta}}{\beta(\beta-1)}, 0<\beta<1, 1<\beta<3. \\ 3(p+2)\ln(p+2) - p(5\ln 5 - 2\ln 2) - 6\ln 2, \beta=1. \end{cases} \tag{6.2}$$

Here the "super exponent" $\beta$ is strictly limited to the interval from zero to three. $a$ is an "intermittency parameter". When it vanishes, there is K41 scaling. $C_3 > 0$ is proportional to the dissipation $\varepsilon$. We do not presently have a value for the proportionality constant. $\ell_o$ is the infrared termination of the scaling laws. $\zeta_3 = 1$, but we have not shown that here. Underlying the scaling laws is the axisymmetric component $\mathcal{J}_o$ of the joint pdf $\mathcal{J}$ for $\chi^+$ and $\chi^-$. The axisymmetric component $\mathcal{J}_o$ is characterized by its radial profile, i.e., $\mathcal{J}_o(x, y, \ell) = P_0(\sqrt{x^2 + y^2}; \ell)$, which is self similar, namely

$$P_0(r;\ell) = C(\ell)f\left(\frac{\ln r - \mu(\ell)}{\sigma(\ell)}\right) = C(\ell)g\left(\left(\frac{r}{e^{\mu(\ell)}}\right)^{1/\sigma(\ell)}\right) \quad . \tag{6.3}$$

We recognize the similarity as affine symmetry in $\ln r$, and $g(r) = f(\ln r)$. The analysis also provided formulas for all items in (6.3). First,

$$\sigma(\ell) = \left(\ln \ell_o - \ln \ell\right)^{1/\beta} \tag{6.4}$$

showing $\ln \ell_o$ as a virtual origin. Then,

$$3\mu(\ell) - \ln\left(\frac{5C_3}{2}\right) - \zeta_3\left(\ln\ell - \ln\ell_o\right) = a\left(\ln\ell_o - \ln\ell\right)\begin{cases} \dfrac{2^{\beta}-5^{\beta}}{\beta(\beta-1)}, 0<\beta<1, 1<\beta<3. \\ 3\ln\left(\ln\ell_o - \ln\ell\right) + 5\ln 5 - 2\ln 2, \beta=1. \end{cases}$$

and

$$C(\ell)\pi\left(\frac{5C_3}{2}\left(\frac{\ell}{\ell_o}\right)^{\zeta_3}\right)^{-2/3} = \begin{cases} \left(\dfrac{\ell}{\ell_o}\right)^{\frac{a}{3\beta(\beta-1)}\left(5\times2^\beta-2\times5^\beta\right)} &,0<\beta<1, 1<\beta<3. \\ \left(\dfrac{\ell}{\ell_o}\right)^{\frac{-2a}{3}\left(5\ln5-8\ln2\right)} &,\beta=1. \end{cases}$$

(6.6)

The similarity profile $g(r)$ is expressed through a new class of special functions $b(r)$ which depends parametrically on $\beta$. Graphs of these functions are shown in Figure 6. We have

$$g(r) = b(r^{\left|\frac{a}{\beta(\beta-1)}\right|}) \ , \quad b(r) = \mathcal{M}^{-1}\left[z^{-1}e^{z^\beta\,\text{sgn}(\beta-1)};r\right] \ \text{for } \beta\neq1$$

(6.7)

and

$$g(r) = |a|^{-1}b(|a|\left(|a|r\right)^{1/|a|}) \ , \quad b(r) = \mathcal{M}^{-1}\left[z^{-1}e^{z\ln z};r\right] \ \text{for } \beta=1 .$$

(6.8)

Formulas (6.1-8) constitute the complete result of the analysis in this paper.

The analysis leaves us with $\zeta_3$, $C_3>0$, $\ell_o, a\leq0$, and $0<\beta<3$. These are not modeling parameters. They are undetermined constants reflecting how far we can go analytically with the combination H1 and H2, but nothing else. By other means, we will elsewhere attempt to establish something akin to the four-fifth law, that is $\zeta_3=1$ and $C_3=Const\cdot\varepsilon$. Moreover, we envision that Navier Stokes equations lead to equations for $a$ and $\beta$. The mathematics behind this vision is only sketchy at this time and may not come to fruition, but if it eventually does then there should be no undetermined parameters left, only the unavoidable normalizations $\ell_o$ and $\varepsilon$. These are intrinsic normalizations for the inertial range: $\ell_o$ is the infrared virtual origin, and $\varepsilon$ is the energy flux. That such a complete analysis without modeling is possible hinges on H1 and H2 being true symmetry properties for equilibrium turbulence in limit of high Reynolds numbers.

Our theoretical attack on the intermittency problem uses a non-traditional set of variables to formulate symmetry. Let us discuss why the choice of variables matters. A good place to start is to quote Frisch (1990, p xi) who wrote: "Modern work on turbulence focuses to a large extent on trying to understand the reasons for the partial failure of the 1941 theory (K41)." Our analysis offers a constructive resolution of this issue. The spirit of K41 calls for a global scaling invariance (i.e., symmetry) in the inertial range because the physics is the same at all scales, namely inertia. That idea comes to fruition in our work, but it requires that turbulence be described in a suitable framework. That is, we need an appropriate set of basis functions and corresponding variables. Only then does symmetry become apparent. K41 thus goes astray technically through its choice of variables. There are two problems with the variables used by K41. First, they define scale inconsistently with the idea of global scaling invariance. Second, they represent a one dimensional Cartesian projection of a two dimensional pdf that should be described in polar coordinates. Let us discuss these two problems in turn.

First, in the tradition of the Karman-Howarth equation for homogeneous turbulence, K41 follows the common practice of using velocity increments in physical space, i.e., $\delta v_\parallel(\ell)$ or $\delta v_\perp(\ell)$. These can be measured experimentally. Corresponding to probes at two points, $P$ and $Q$, the implied definition of scale is $\ell = \left|\overrightarrow{PQ}\right|$. By finding velocity increments we filter $\vec{u}$. Unfortunately, the filter has poor three dimensional characteristics. This fact becomes clear when we think a plane wave with wave vector $\vec{k}$. The filter considers only the component of $\vec{k}$ along $\overrightarrow{PQ}$. So a low frequency filter (i.e., large $\ell = \left|\overrightarrow{PQ}\right|$) picks up high frequency waves as illustrated in Figure 8. The implication is well known for the energy spectrum in isotropic turbulence; see Hinze (1987, p 209). In fact, there is a way to convert the one dimensional energy spectrum $E_1\left(k_1\right)$ obtained from $PQ$-probe into a three dimensional spectrum $E\left(\left|\vec{k}\right|\right)$. The energy spectra arise from second order structure functions and afore mentioned conversion formula is special to that case. For structure functions of arbitrary order, we lack corresponding formulas. It is important to understand that this whole issue disappears when we operate in Fourier space from the very beginning for it is well known from the wavelet literature that Fourier space provides the ideal separation into scales; e.g., Strang and Nguyen (1996).

Second, even when the variables are defined in Fourier space, the symmetry is still not obvious. Shell models illustrate this point. For example, the popular shell model "GOY" has been studied extensive because it exhibits anomalous scaling and inertial range intermittency in a simpler setting than the Navier Stokes equations (e.g., Frisch 1990 p 176, Jensen et al 1994 , Biferale 2003). Like Zimin's model, GOY has two real variables in each shell. Conceptually, we have the same situation for both the shell model and the Navier Stokes in $(\chi^-, \chi^+)$-variables. That is, a two dimensional pdf describes the shell statistics. Its natural description is through azimuthal modes (2.5), where the mean radial profile $P_0(r, \ell)$ is the primary focus.  As described in Section 2, $P_0(r, \ell)$ can scale nicely through affine symmetry in the logarithmic coordinate $\Omega = \ln r$.  It is as simple as resetting the origin and the unit on the $\Omega-$axis (see Figures 3 and 4).  The usual statistics, gathered when working with shell models, include the pdf's for the amplitude $\left( \left( \chi^- \right)^2 + \left( \chi^+ \right)^2 \right)^{1/2}$ together with individual components $\chi^-$ and $\chi^+$.  Those statistics hide our affine symmetry.  In particular, we see only the usual intermittency and anomalous scaling in the pdf's for $\chi^-$ and $\chi^+$.  That is, they look similar to the pdf for $\delta v_{\parallel}$; see figure 2b and Biferale (2003 p 448).  By way of an analytic example, Melander and Fabijonas (2006) illustrated how the pdf's for $\chi^-$ and $\chi^+$ can hide our affine symmetry.  In that example, we used the explicit inverse Mellin transform for $\beta = 2$ to avoid the technicalities of Sections 3-5.  Our example was constructed from affine symmetry so as to have scaling exponents matching those in Kolmogorov (1962) yet the pdf's for $\chi^-$ and $\chi^+$ looked like those for the usual velocity increments.  That analytic example makes the point that intermittency and self-similarity can coexist. However, neither collapse in Figure 4 corresponds to $\beta = 2$.  So $\beta = 2$ is not the value selected by Navier Stokes' or Zimin's shell model.

The benefit of using good variables is that the underlying symmetry becomes apparent and the analysis manageable. On the basis of the good variables we have formulated a broad new framework for analyzing turbulence. With appropriate modifications the framework can be applied in related situations, e.g., 2D-turbulence, passive scalars, and infrared cascades.  We must, however, point out one important caveat. That is, other than intermittency and anomalous scaling, we have little to compare directly with the traditional approach based on velocity increments. Presently, no direct comparison with experimental data is possible. One could have hoped that at our scaling exponents would match $\zeta_{\|p}$ or $\zeta_{\perp p}$, but the DNS data shows that they do not. Thus, presently, we can only compare our theoretical results with DNS data processed according to our formulas. Elsewhere, we will report a more detailed analysis of the present data, including the techniques for finding the constants. The thrust of this paper is theoretical; data is included only to back up the hypothesis H1 and H2.

**Acknowledgements**. The author is grateful for helpful discussions with B. Fabijonas, who also produced the graphs in Figure 6. Moreover, the DNS data provided by the LANL group D. Holm. S. Kurien and M. Taylor is much appreciated.

# 8. Appendix I

Let us reconsider the choice of the non-linear transformation $\Omega(r)$ in H1. While a logarithmic function is an obvious way of mapping the semi-infinite axis onto the full axis, one could reasonably ask if it is meaningful to use another function in H1. The criterion as to what constitutes a meaningful $\Omega(r)$ will have to be that the transformation must not ruin statistical self-similarity (K41-scaling) by mapping it onto something not self similar in $\Omega(r)$. That is, if $P_0(r;\ell)$ possesses K41-scaling using the *r*-variable then it should also scale normally in $\Omega(r)$. With that restriction, we will show that only logarithmic functions can be used as $\Omega(r)$ in H1.

Our constraint that a function $F(r,\ell)$ obeying self-statistical similarity should also scale normally in $\Omega$ requires that we simultaneously have

$$F(r,\ell) = f_1(r/s(\ell)) \tag{8.1}$$

and

$$F(r,\ell) = f_2((\Omega(r) - \mu(\ell))/\sigma(\ell)), \tag{8.2}$$

where $f_1, f_1, s, \mu, \sigma$ are unspecified. By defining $\Phi \equiv f_2^{-1} \circ f_1$ we can rewrite (8.2) as

$$\Phi\left(\frac{r}{s(\ell)}\right) = \frac{\Omega(r) - \mu(\ell)}{\sigma(\ell)}. \tag{8.3}$$

Now we pick an arbitrary constant $C$ in the range of $s$ and let $\ell_C$ denote the solution to $s(\ell) = C$. Substituting $\ell_C$ and $x = r/C$ into (8.3), we obtain

$$\Phi(x) = \frac{\Omega(Cx) - \mu(\ell_C)}{\sigma(\ell_C)} \tag{8.4}$$

Using (8.4) on the left hand side of (8.3), we eliminate $\Phi$ and obtain:

$$\frac{\Omega(Cr/s(\ell)) - \mu(\ell_C)}{\sigma(\ell_C)} = \frac{\Omega(r) - \mu(\ell)}{\sigma(\ell)}. \tag{8.5}$$

We can rewrite (8.5) as

$$\Omega(r/s_1(\ell)) = \frac{\Omega(r) - \mu_1(\ell)}{\sigma_1(\ell)} \tag{8.6}$$

by defining $s_1(\ell) \equiv s(\ell)/C$, $\sigma_1(\ell) \equiv \sigma(\ell)/\sigma(\ell_C)$, and $\mu_1(\ell) \equiv \mu(\ell) - \mu(\ell_C)\sigma_1(\ell_C)$.

Equation (8.6) states that $\Omega$ maps self-statistical similarity to normal scaling. At least locally, $\mu_1$ is an invertible function of $\ell$. Thus, $s_1$ and $\sigma_1$ are implicit functions of $\mu_1$, i.e., $s_1(\ell) = s_2(\mu_1(\ell))$ and $\sigma_1(\ell) = \sigma_2(\mu_1(\ell))$. This allows us to eliminate $\ell$ from (8.6):

$$\Omega(r/s_2(x)) = \frac{\Omega(r) - x}{\sigma_2(x)}, \tag{8.7}$$

which is a functional equation for $\Omega, s_2$ and $\sigma_2$. We solve it by reformulating it into one whose solution is known. That requires some additional transformations. First, we define $w \equiv \ln r$, $\tau \equiv -\ln(s_2(x))$, and $A \equiv \rho \circ \exp$, so that $\Omega(r) = \Omega(\exp(\ln r)) = A(w)$ and $\Omega(r/s_2(x)) = \Omega(\exp(\ln(r/s_2(x)))) = A(w + \tau)$. Next we need to express $1/\sigma_2(x)$ and $-x/\sigma_2(x)$ as functions of $\tau$. Noting that $x = s_2^{-1}(e^{-\tau})$, we have:

$$B(\tau) \equiv \frac{1}{\sigma_2(x)} = \frac{1}{\sigma_2(s_2^{-1}(e^{-\tau}))}, \tag{8.8}$$

and

$$D(\tau) \equiv -\frac{x}{\sigma_2(x)} = -\frac{s_2^{-1}(e^{-\tau})}{\sigma_2(s_2^{-1}(e^{-\tau}))}. \tag{8.9}$$

Using these expression, we bring (8.7) into the form

$$A(w + \tau) = A(w)B(\tau) + D(\tau) \tag{8.10}$$

which is covered by theorem 1 on p.150 in Azcel (1966). Moreover, $\Omega$, and hence $A$, is a monotone function so the corollary on page 150 applies. It states that (8.10) has exactly two solution systems:

$$A(w) = \gamma w + \alpha, \quad B(\tau) = 1, \quad D(\tau) = \gamma\tau \tag{8.11}$$

and

$$A(w) = \gamma e^{cw} + \alpha \ , \quad B(\tau) = e^{c\tau} \ , \quad D(\tau) = \alpha(1 - e^{c\tau}) \qquad (8.12)$$

where $c \neq 0$, $\gamma \neq 0$, and $\alpha$ are arbitrary constants. The first system, (8.11), yields $\Omega(r) = A(\ln r) = \gamma \ln(r/r_o)$ with $\alpha = -\gamma \ln r_o$. These are the logarithmic functions. The constants $\alpha, \gamma$ are inconsequential. This system is precisely the one represented by $\ln r$ in H1.

The second system yields $\Omega(r) = A(\ln r) = \gamma r^c + \alpha$. The corresponding structure functions for $P_0(r, \ell)$ are

$$S_p(\ell) = \tilde{C}(\ell) \int_0^\infty r^{p+1} f \left( \frac{\Omega(r) - \mu(\ell)}{\sigma(\ell)} \right) dr = \tilde{C}(\ell) \int_0^\infty r^{p+1} f \left( \frac{\gamma r^c + \alpha - \mu(\ell)}{\sigma(\ell)} \right) dr \quad (8.13)$$

where $\tilde{C}(\ell)$ is determined by the requirement $S_o(\ell) \equiv 1$ and $f$ is unknown. Let us examine how (8.13) fares when subjected to H2. The question is which functions $f(x), \mu(\ell), \sigma(\ell)$ and constants $\alpha, \gamma, c$ admit power laws $S_p(\ell) = C_p \ell^{\zeta_p}$. (Note we can not use the latter two equations in (8.12) because they only apply when there is statistical self-similarity in $r$.) The first step is to change variables in the integral. Let $\kappa(\ell) \equiv (\gamma/\sigma)^{1/c}$ and $\eta(\ell) \equiv (\mu - \alpha)/\sigma$, the substitution $u = \kappa r$ then yields

$$S_p(\ell) = C_p \ell^{\zeta_p} = \tilde{C} \kappa^{-(p+2)} I_p \ , \qquad (8.14)$$

where

$$I_p(\ell) \equiv \int_0^\infty u^{p+1} f \left( u - \eta(\ell) \right) du \ . \qquad (8.15)$$

Taking the logarithm on both sides of (8.14) yields

$$\ln C_p + \zeta_p \ln \ell = \ln \tilde{C}(\ell) - (p+2) \ln \kappa + \ln I_p \ . \qquad (8.16)$$

Differentiating on both sides with respect to $\ln \ell$ yields

$$\zeta_p = K(\ell) - (p+2)Q(\ell) + \frac{\ell}{I_p}\frac{\partial I_p}{\partial \ell}, \tag{8.17}$$

where $K(\ell) = \ell\partial(\ln \tilde{C})/\partial\ell$ and $Q(\ell) = \ell\partial(\ln \kappa)/\partial\ell$. Using integration by parts we have

$$\frac{\partial I_p}{\partial \ell} = -\eta'(\ell)\int_0^\infty u^{p+1}f'(u-\eta(\ell))du = (p+1)\eta'(\ell)\int_0^\infty u^p f(u-\eta)du = (p+1)\eta'(\ell)I_{p-1}. \tag{8.18}$$

We can eliminate $I_p$ from (8.17) with the help of the power laws (8.14), that is

$$\frac{\ell}{I_p}\frac{\partial I_p}{\partial \ell} = (p+1)\ell\eta'\frac{I_{p-1}}{I_p} = \kappa^{-1}(p+1)\ell\eta'\frac{\kappa^{p+1}\tilde{C}I_{p-1}}{\kappa^{p+2}\tilde{C}I_p}$$

$$= \kappa^{-1}(p+1)\ell\eta'\frac{S_{p-1}(\ell)}{S_p(\ell)} = \kappa^{-1}(p+1)\ell\eta'\frac{C_{p-1}}{C_p}\ell^{\zeta_{p-1}-\zeta_p} \tag{8.19}$$

Thus, (8.17) takes the form

$$\zeta_p = K(\ell) - (p+2)Q(\ell) + \kappa^{-1}\ell\eta'(p+1)\frac{C_{p-1}}{C_p}\ell^{\zeta_{p-1}-\zeta_p}. \tag{8.20}$$

Recalling that $C_0 = 1$ and $\zeta_0 = 0$, we set $p = 0$ in (8.20) and solve for $K(\ell)$

$$K(\ell) = 2Q(\ell) - \kappa^{-1}\ell\eta'C_{-1}\ell^{\zeta_{-1}}, \tag{8.21}$$

so that (8.20) can be rewritten as

$$\zeta_p = -pQ(\ell) + \kappa^{-1}\ell\eta'\left[(p+1)\frac{C_{p-1}}{C_p}\ell^{\zeta_{p-1}-\zeta_p} - C_{-1}\ell^{\zeta_{-1}}\right]. \tag{8.22}$$

Setting $p = 1$ allows us to find $Q(\ell)$:

$$Q(\ell) = -\zeta_1 + \left(\kappa^{-1}\ell\eta'\right)\left[\frac{2}{C_1}\ell^{-\zeta_1} - C_{-1}\ell^{\zeta_{-1}}\right]. \tag{8.23}$$

Thereby, we can write (8.22) as

$$\zeta_p - p\zeta_1 = \left(\kappa^{-1}\ell\,\eta'\right)\left[(p+1)\frac{C_{p-1}}{C_p}\ell^{\zeta_{p-1}-\zeta_p} + (p-1)C_{-1}\ell^{\zeta_{-1}} - \frac{2p}{C_1}\ell^{-\zeta_1}\right]. \quad (8.24)$$

We eliminate $\kappa^{-1}\ell\,\eta'$ by considering the ratio

$$\frac{\zeta_p - p\zeta_1}{\zeta_q - q\zeta_1} = \frac{(p+1)\dfrac{C_1 C_{p-1}}{C_p}\ell^{\zeta_{p-1}-\zeta_p+\zeta_1} - (p-1)C_1 C_{-1}\ell^{\zeta_{-1}+\zeta_1} - 2p}{(q+1)\dfrac{C_1 C_{q-1}}{C_q}\ell^{\zeta_{q-1}-\zeta_q+\zeta_1} - (q-1)C_1 C_{-1}\ell^{\zeta_{-1}+\zeta_1} - 2q} \quad . \quad (8.25)$$

In order for the right hand side to be independent of $\ell$, we must have $\zeta_{p-1} - \zeta_p + \zeta_1$ independent of $p$. Thus,

$$\zeta_p = Cp + \varpi(p), \qquad \varpi(p+1) = \varpi(p) \quad (8.26)$$

where $\varpi$ is a periodic function with period one. From (1.9) we know that $\zeta_p$ must be concave down in ultraviolet regime, $\ell \to 0$. Thus, $\varpi \equiv 0$ and only $\zeta_p = Cp$ is possible in (8.12). That leaves only (8.11) for the intermittency. Hence, we have $\Omega = \ln r$ as the only choice in H1.

# 9. Appendix II

## Special case $\beta = 0$

Integrating (4.8) once we find an equation different from (4.11), namely:

$$\chi^2 \eta''(\chi) = ab\ln\chi + c_1. \quad (9.1)$$

We readily obtain

$$\eta(\chi) = -\frac{ab}{2}\left(\ln\chi\right)^2 - (ab + c_1)\ln\chi + c_2\chi + c_3 \quad (9.2)$$

where $c_2 \chi + c_3$ is the homogeneous solution to (3.21). Because the homogeneous solution does not affect $S_p(\ell)$, we can, without loss of generality, take $c_2 = c_3 = 0$. Using (3.18), we find

$$G(z) = z^{-(ab+c_1)} \exp\left(-\frac{ab}{2}(\ln z)^2\right).$$  (9.3)

Again, we need to determine $c_1$ so that $g(0) = \mathscr{M}^{-1}[G(z);0]$ exists and is finite. We proceed as in Section 5.2, expect this time we can use $g$ and $G$ instead of $\flat$ and $\mathscr{B}$. With $\tilde{g}(r,\delta)$ defined from $G(z)$ as in (5.14), the steps (5.15-18) lead us to

$$\tilde{g}(0,\delta) = \delta G(\delta) = \delta^{1-(ab+c_1)} e^{-\frac{ab}{2}(\ln \delta)^2}.$$  (9.4)

Because of the last factor in (9.4), we can only have a finite non-zero limit as $\delta \to 0^+$ provided that $ab = 0$. From the first factor, we see that the same requirement implies $c_1 = 1$. However, $ab = 0$ gives us one of the trivial solutions to (3.21) and thus K41 scaling rather than intermittency. Therefore, we conclude $\beta = 0$ is not permissible.

## Special case $\beta = 1$

This special case follows the main case through (4.17), so the formulae for both $B(p)$ and $\sigma(\ell)$ remain unchanged. Again, we are allowed to take $b/\beta = -1$, so $b = -1$. However, we get new expressions for $A(p)$ and $\eta(\chi)$. Integrating (4.14) together with (4.17) yields

$$A(p) = \frac{a}{3}\Big(3(p+2)\ln(p+2) - (p+2)(5\ln 5 - 2\ln 2) - 10\ln(2/5)\Big).$$  (9.6)

Moreover, by integrating (4.11) we obtain

$$\eta(\chi) = -a\chi\ln\chi - c_1\ln\chi + c_2\chi + c_3. \tag{9.7}$$

As before, we can take $c_2 = c_3 = 0$. Thereby, we have

$$G(z) = z^{-c_1}e^{-az\ln z}. \tag{9.8}$$

Again, we need to determine $c_1$ so that $g(0) = \mathcal{M}^{-1}[G(z);0]$ exists, is finite and positive. Proceeding as for $\beta = 0$, we obtain

$$\tilde{g}(0,\delta) = \delta G(\delta) = \delta^{1-c_1}e^{-a\delta\ln\delta} \tag{9.9}$$

As $\delta \to 0^+$, we have a finite positive limit precisely when $c_1 = 1$. Thus,

$$G(z) = z^{-1}e^{-az\ln z} \Rightarrow g(r) = \frac{1}{2\pi i}\int_{c-i\infty}^{c+i\infty} z^{-1}e^{-az\ln z - z\ln r}dz, \qquad c > 0. \tag{9.10}$$

The next step is to check that the integral converges. Along the line $z = c + iy$, we have, using $y = \rho\sin\theta$,

$$R = \mathrm{Re}\left(-az\ln z - z\ln r\right) = -ac\ln\rho + a\rho\theta\sin\theta - c\ln r. \tag{9.11}$$

Since $\theta \to \pm\pi/2$ as $y \to \pm\infty$ along the line we have the asymptotic expression

$$R \sim -ac\ln\rho + a\rho\pi/2 - c\ln r. \tag{9.12}$$

The leading order is $R \sim a\rho\pi/2$. Consequently, the integral (9.10) converges when $a < 0$ and diverges when $a > 0$. ($a = 0$ is not an option, because that results in a trivial solution to (3.22) and thus no intermittency.) Since $a < 0$, we can express (9.10) in normal form by defining two new functions:

$$\mathcal{B}(z) \equiv z^{-1}e^{z\ln z} \tag{9.13}$$

and

$$b(r) \equiv \mathcal{M}^{-1}[\mathcal{B}(z);r], \tag{9.14}$$

Thereby,

$$G(z) = z^{-1} e^{|a||z \ln z|} = |a|^{-(|a|+1)z} \mathcal{B}\left(|a|\,\big|\,z\right).$$  (9.15)

and

$$g(r) = \mathcal{M}\left[\,|a|^{-(|a|+1)z}\,\mathcal{B}\left(|a|\,\big|\,z\right); r\right] = |a|^{-1}\,b\left(|a|\left(|a|\,r\right)^{|a|^{-1}}\right).$$  (9.16)

The contour integral for $b(r)$ is (9.10) with $a = -1$. Finally, we need to check if $b(r) \geq 0$. To do so, we look for the curve describing the zero phase. That curve is defined by the equation:

$$\mathrm{Im}\left(z \ln z - z \ln r\right) = \rho\theta\cos\theta + \left(\sin\theta\right)\rho\ln\rho + \left(\sin\theta\right)\rho\ln r = 0.$$  (9.17)

Rewriting the expression, we find that the steepest decent contour is given by

$$\mathcal{S}: \quad \rho(\theta) = \begin{cases} r^{-1} e^{-\theta\cot\theta}, \theta \neq 0 \\ r^{-1} e^{-1}, \theta = 0 \end{cases}.$$  (9.18)

The curve is shown in Figure 7f. Along this curve, we have

$$\mathrm{Re}\left(z \ln z - z \ln r\right) = -\rho\theta\sin\theta + \left(\cos\theta\right)\rho\ln\rho - \left(\cos\theta\right)\rho\ln r.$$  (9.19)

For large $\rho$, the term $\left(\cos\theta\right)\rho\ln\rho$ dominates. Since $\mathcal{S}$ approaches infinity through the left half plane we have $\cos\theta < 0$ asymptotically. The integrand in (9.10) thus decays faster than algebraically; consequently, we can deform the integration path to $\mathcal{S}$. Using the fact that $\rho(\theta)$ is an even function, we find

$$b(r) = \frac{1}{\pi}\int_0^\pi e^{(\ln\rho)\rho\cos\theta - \rho\theta\sin\theta - (\ln r)\rho\cos\theta} d\theta.$$  (9.20)

Here the integrand is positive, so $b(r) > 0$. Thus, $\beta = 1$ is permissible. The graph of $b(r)$ is shown in Figure 6b.

## 10. Figure Captions

Figure 1.  Phase portrait where symmetry exists only on the attractor.

Figure 2.  Typical picture of intermittency: pdf's for different $\ell$ fail to collapse when plotted in units of standard deviations.  (a) $\delta v_{\parallel}\left(\ell\right)$-pdf  from  $512^3$ DNS obtained from LANL-group Holm, Kurien & Taylor.  (b) Shell model data:  the random variable is $x = \chi^+ + \chi^-$,  one of the two real shell variables in Zimin's shell model.

Figure 3.  Schematic showing an axisymmetric monopole before (a) and after  (b) coordinate transformation from a semi-infinite to a full axis.  $P_0(r)$ need not be flat at $r = 0$,  just  $0 < P_0(0) < \infty$.  After the transformation there are horizontal asymptotes as $\Omega \to \pm\infty$.

Figure 4.  The radial profile $P_0(r, \ell)$ for various $\ell$ and the collapse.  (a) Same shell model data as in figure 2; $\ell = 2^{-n}$, where $n$ is indicated in the figure. The bottom graphs show $P_0(r, \ell)$ as computed from the data. The top graph shows the collapse obtained by abscissa scaling plus vertical and horizontal shifts. (b) Same $512^3$ DNS data as in figure 2; $\ell = 2\pi / k$, where is $k$ shown in the figure.

Figure 5. Power law behavior in the inertial range. (a) Shell model data. The power laws are the straight lines in the log-log plot. The data also reveals the virtual origin. The theoretical formula for $C_p$ is used to generated the plot, hence the factor $(p+2)/2$ in front of $S_p(\ell)$. (b) $512^3$-DNS data plotted in ESS format.

Figure 6. Graphs of the function $b(r)$ in the permissible interval $0 < \beta < 3$: (a) $0 < \beta < 1$, note the compact support; (b) $\beta = 1$; (c) $1 < \beta \leq 2$, the solid for $\beta = 2$ is the complementary error function; (d) $2 < \beta < 3$.

Figure 7: Integration paths used for the inverse Mellin transforms follow the zero phase contours in the complex plane. The original integration path is the vertical line. The shaded regions indicate where the integrand diverges at infinity. (a) $0 < \beta < 1$ and $r < 1$; (b) $1 < \beta < 3$ and $r > 1$; (c) $1 < \beta < 2$ and $r < 1$; (d) $2 < \beta < 3$ and $r < 1$. (e) $\beta > 5$; (f) $\beta = 1$.

Figure 8. Sketch showing how a one dimensional probe $PQ$ sees plane waves. In particular, it picks up signals from high frequency waves that hit at an angle.

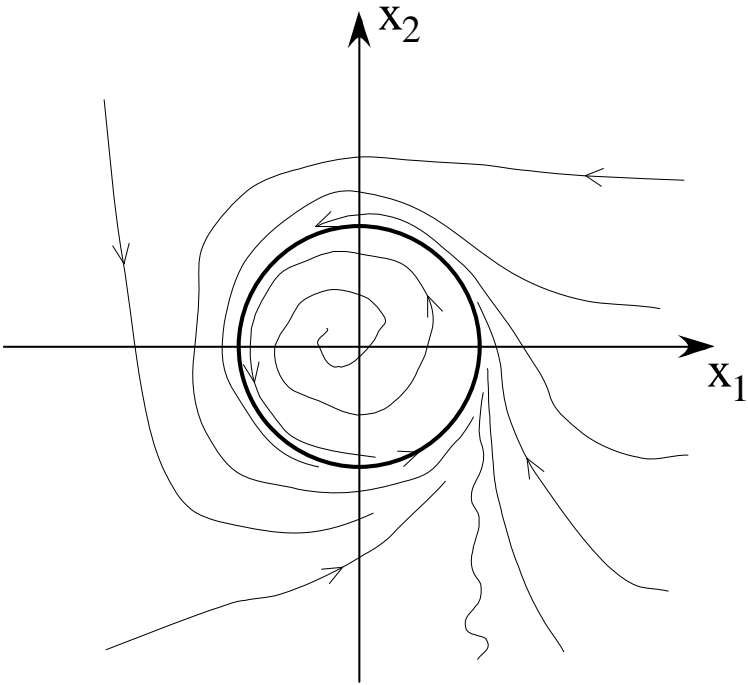

Figure 1

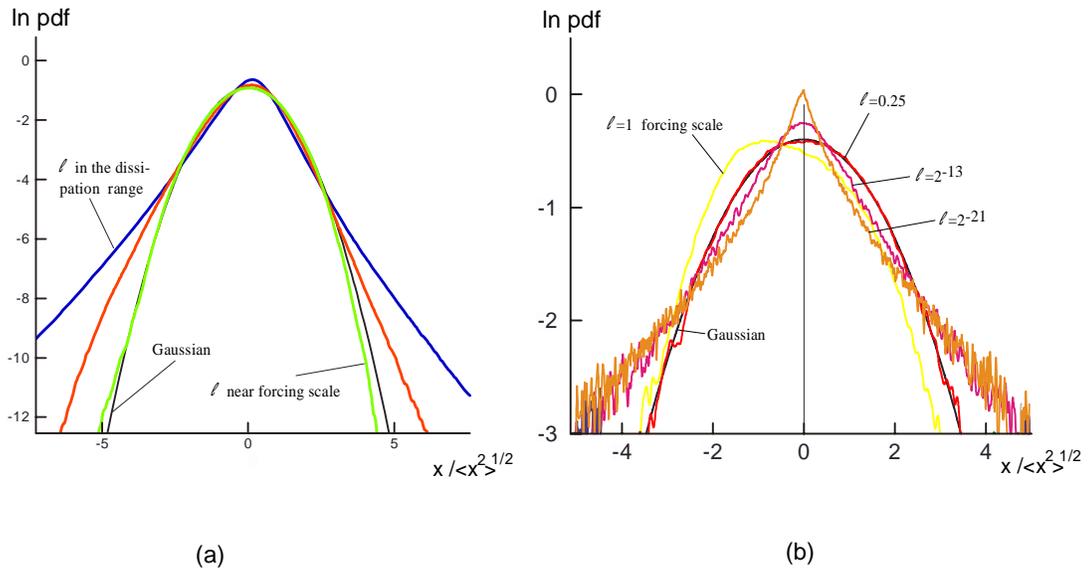

(a)                                                      (b)

Figure 2

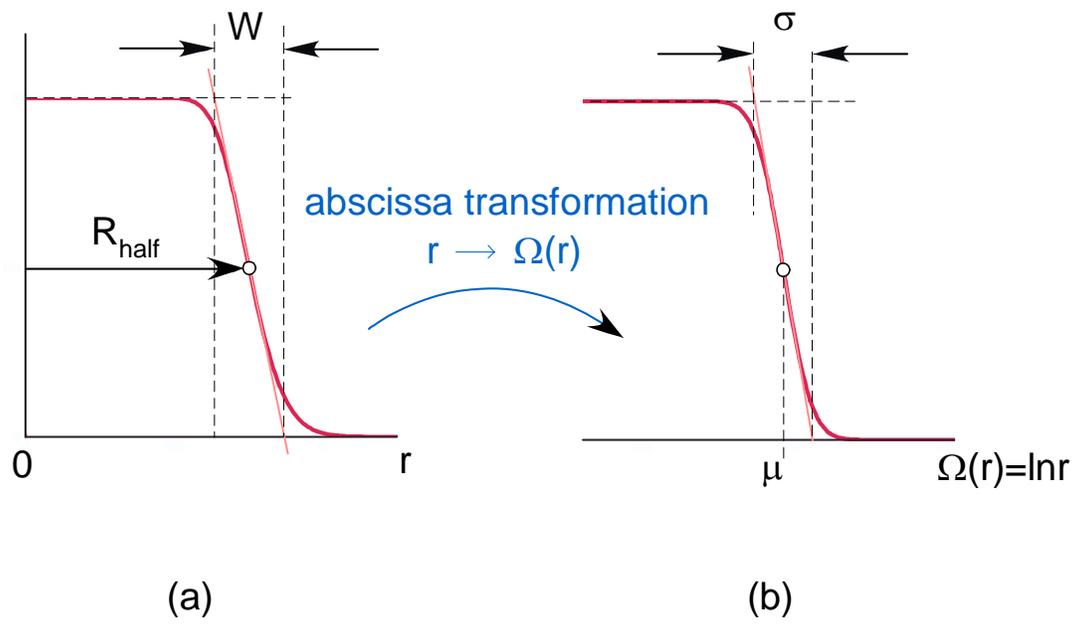

(a)

(b)

Figure 3

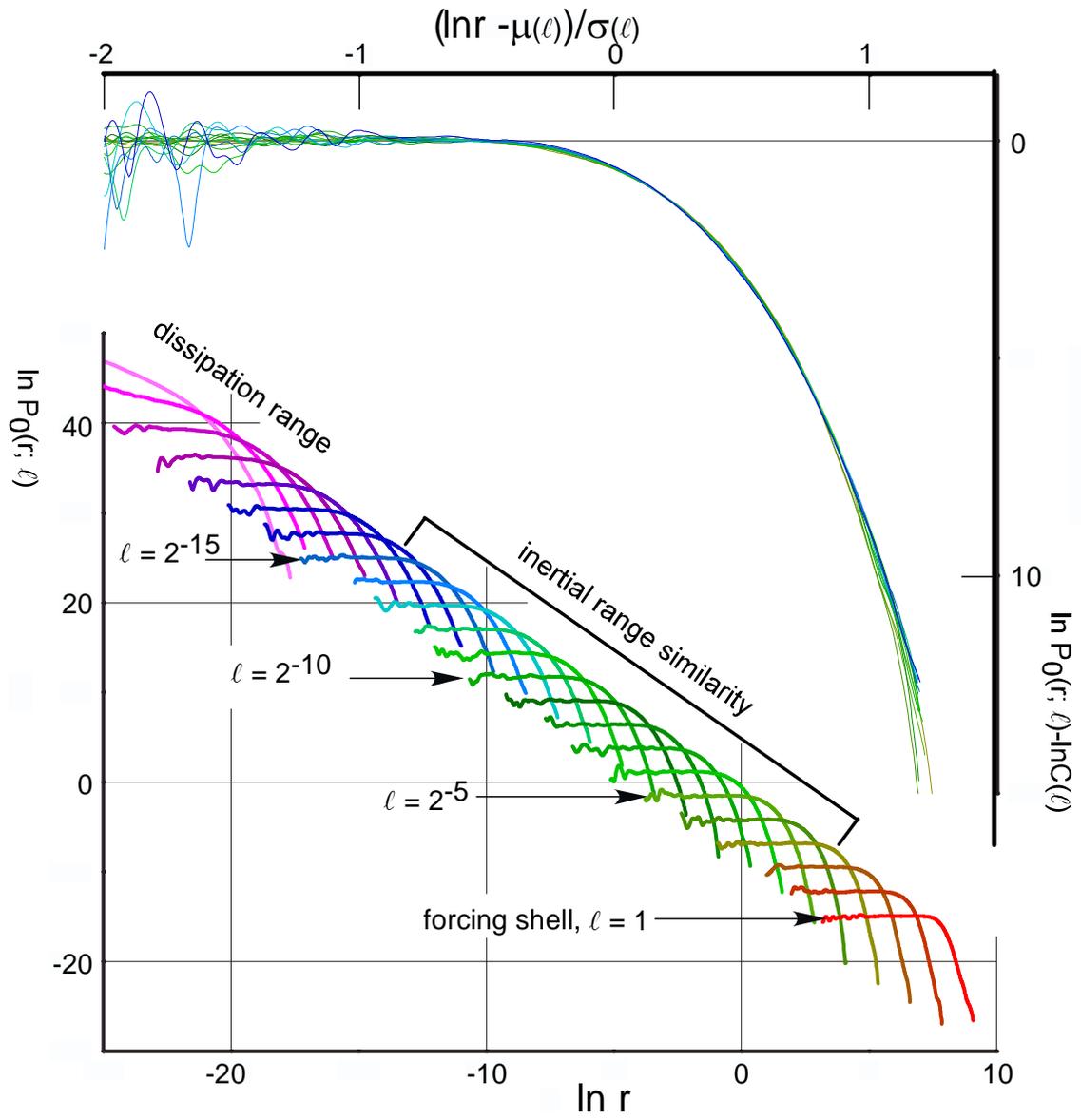

Figure 4a

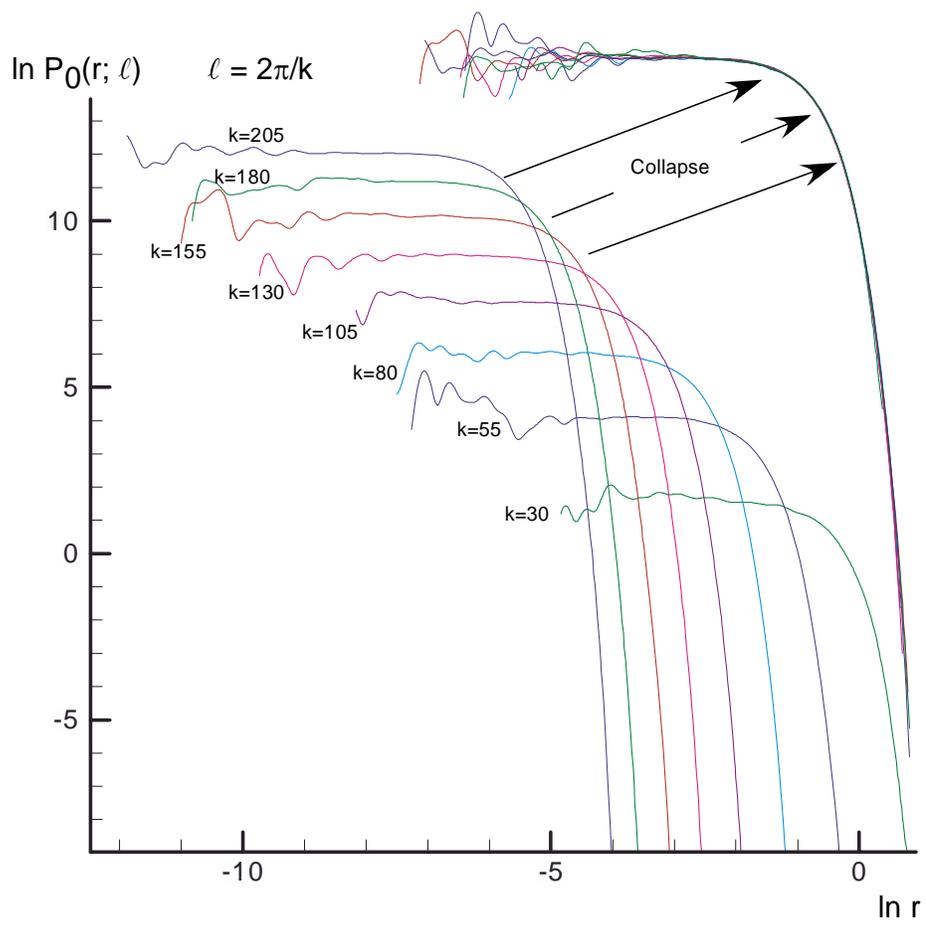

Figure 4b

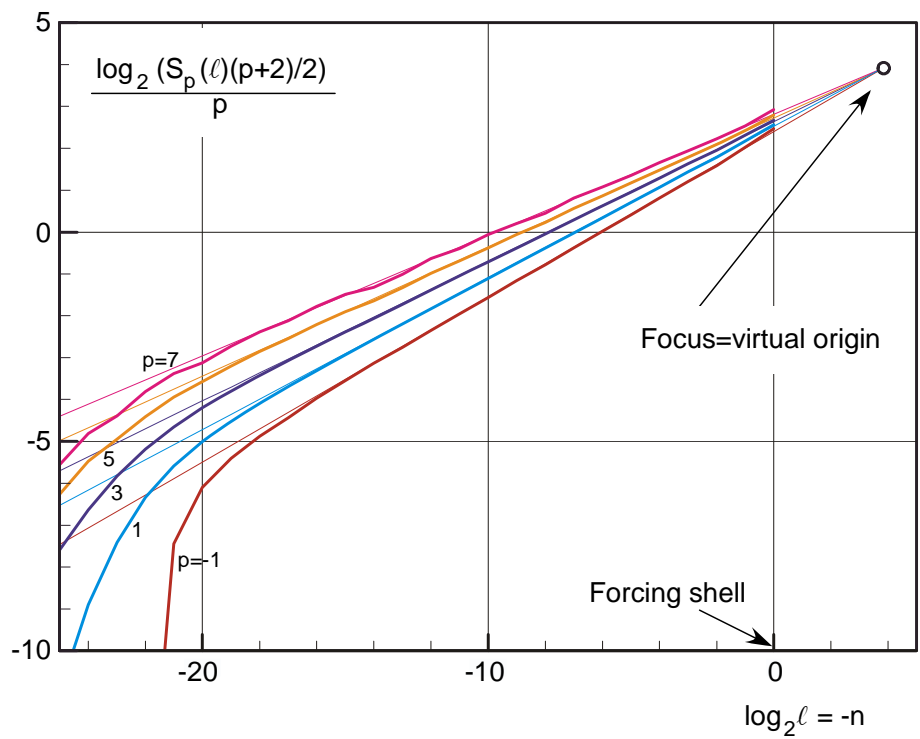

Figure 5a

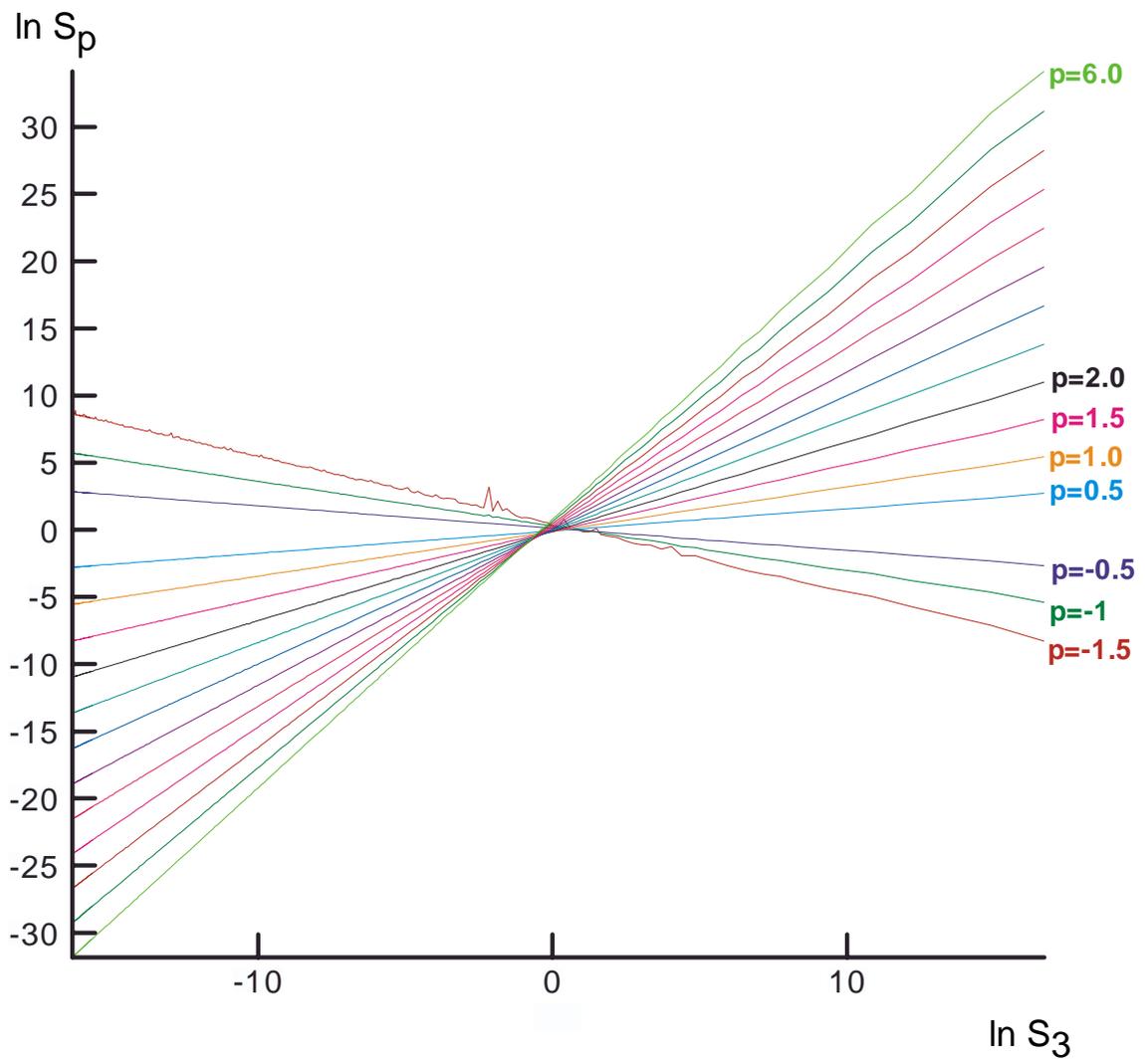

Figure 5b

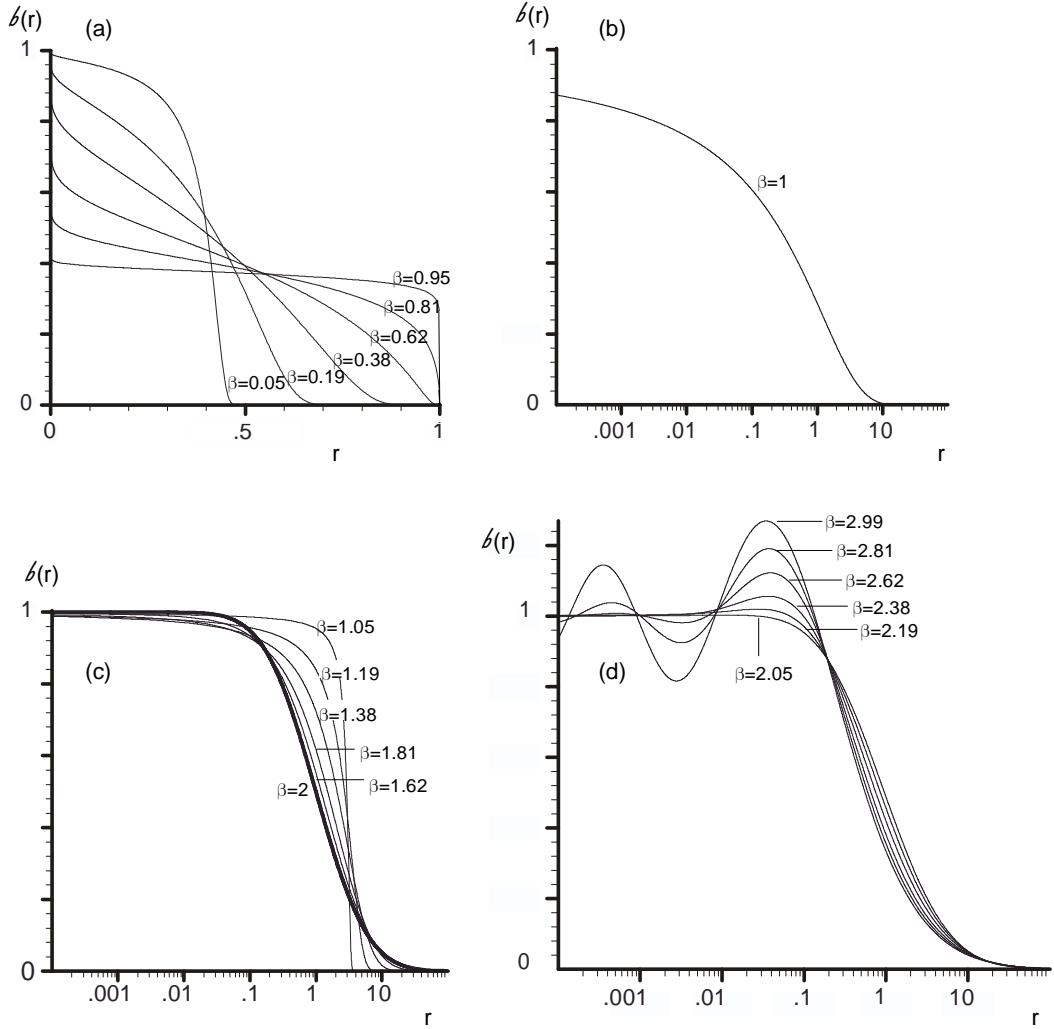

Figure 6

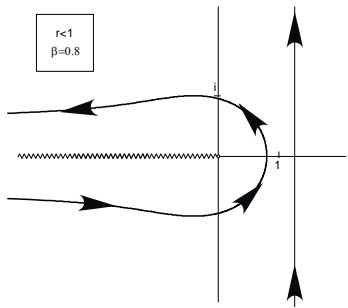

(a)

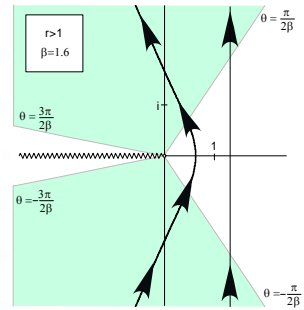

(b)

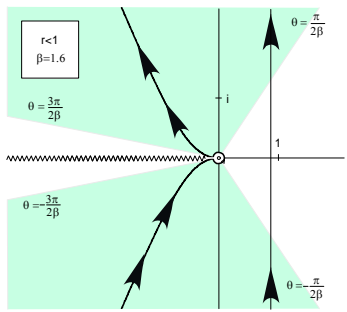

(c)

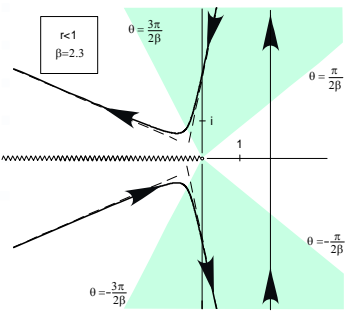

(d)

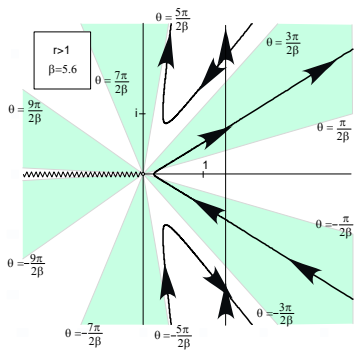

(e)

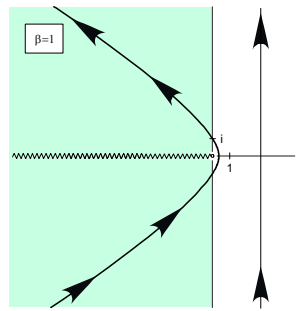

(f)

Figure 7

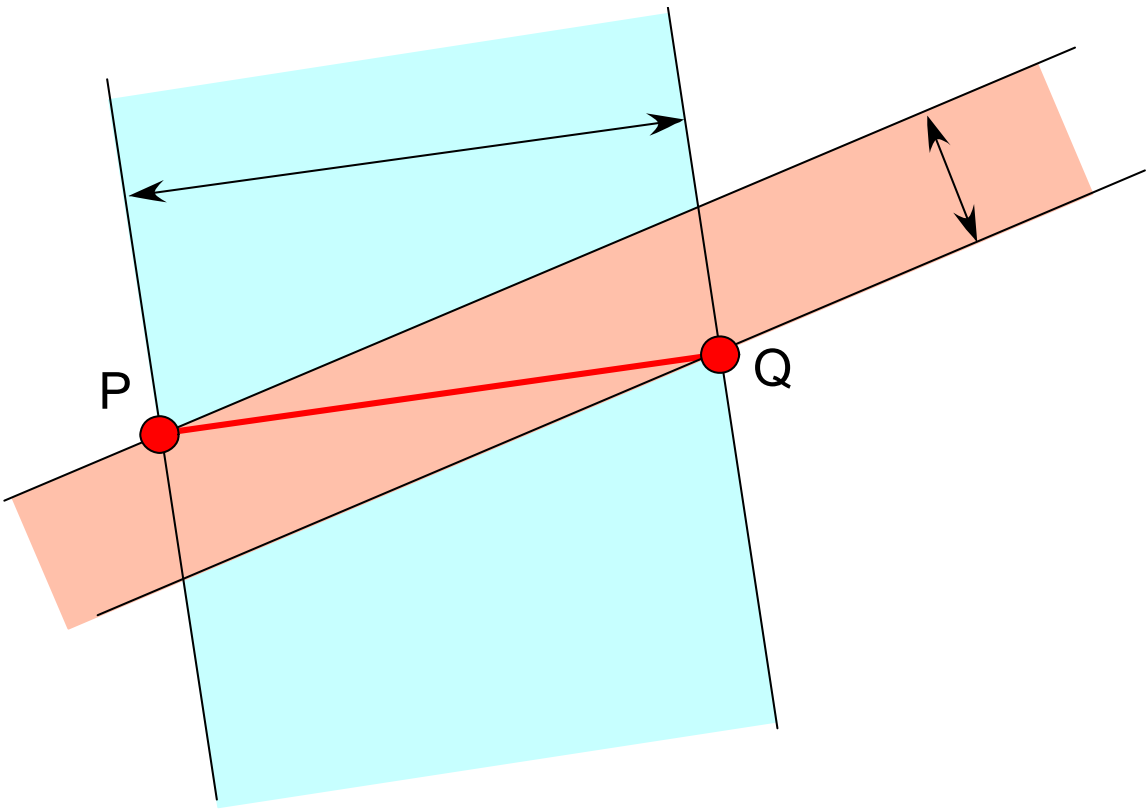

Figure 8